\documentclass[11pt]{article}

\usepackage[T1]{fontenc}
\usepackage{lmodern,microtype}
\usepackage[left=1.15in,right=1.15in,top=1.3in,bottom=1.3in]{geometry}
\usepackage{xcolor}
\usepackage[most,breakable]{tcolorbox}
\usepackage{authblk}

\usepackage{amsmath,amssymb,amsthm,mathtools}
\usepackage[colorlinks=true,linkcolor=blue!45!black,citecolor=blue!45!black,urlcolor=blue!45!black]{hyperref}

\newtheorem{theorem}{Theorem}

\newtheorem{proposition}[theorem]{Proposition}
\newtheorem{lemma}[theorem]{Lemma}
\newtheorem{corollary}[theorem]{Corollary}

\definecolor{mainresultaccent}{HTML}{176B87}
\definecolor{technicalaccent}{HTML}{657A45}
\tcbset{resultbox/.style={
 enhanced, unbreakable, boxrule=0.5pt, arc=1.5pt, outer arc=1.5pt,
 left=5.75pt, right=5.75pt, top=4pt, bottom=4pt,
 grow sidewards by=10pt, before skip=8pt, after skip=8pt
}}
\tcolorboxenvironment{theorem}{resultbox,
 colback=mainresultaccent!9!white, colframe=mainresultaccent!38!white}
\tcolorboxenvironment{lemma}{resultbox,
 colback=technicalaccent!6!white, colframe=technicalaccent!32!white}

\tcolorboxenvironment{proposition}{resultbox,
 colback=technicalaccent!6!white, colframe=technicalaccent!32!white}
\tcolorboxenvironment{corollary}{resultbox,
 colback=mainresultaccent!9!white, colframe=mainresultaccent!38!white}
\allowdisplaybreaks[1]
\newcommand{\cH}{\mathcal{H}}

\newcommand{\Ftest}{\mathcal{F}}

\newcommand{\opnorm}[1]{\left\lVert #1\right\rVert}
\newcommand{\ket}[1]{\lvert #1\rangle}
\newcommand{\bra}[1]{\langle #1\rvert}
\newcommand{\ketbra}[2]{|#1\rangle\!\langle#2|}
\newcommand{\braket}[2]{\langle#1|#2\rangle }

\title{Norm Bounds on Sums of Positive Operators with Applications to Monogamy Games and Quantum Key Distribution}
\author[1,2]{Salman Beigi}
\author[3,2]{Marco Tomamichel}
\affil[1]{\textit{School of Mathematics, Institute for Research in Fundamental Sciences (IPM), P.O. Box 19395-5746, Tehran, Iran}}
\affil[2]{\textit{Centre for Quantum Technologies, National University of Singapore, Singapore 117543, Singapore}}
\affil[3]{\textit{Department of Electrical and Computer Engineering, National University of Singapore, Singapore 117583, Singapore}}
\date{}

\begin{document}

\maketitle

\begin{abstract}
Bounding the acceptance probabilities of quantum measurements
is a recurring task in quantum information theory. An acceptance probability
averaged over a random measurement setting can be bounded uniformly
over the input state by the operator norm of an average
of positive semidefinite operators. We prove a general inequality
for such norms, valid for arbitrary finite families of positive
semidefinite operators and averaging distributions. 
We give two applications of our operator-norm inequality. First, we determine the sharp robustness
threshold for the parallel BB84 monogamy-of-entanglement game.
For error thresholds $\beta,\gamma\in[0,1/2]$, when the two players
may make at most $\beta n$ and $\gamma n$ errors, respectively, the optimal winning probability decays
exponentially if $(1-2\beta)^2+(1-2\gamma)^2>1$, tends to one
exponentially fast under the reverse strict inequality, and remains
bounded away from zero on the boundary. Second, we establish the achievability of every
asymptotic key rate $0<R<1-2h_2(\gamma)$ for one-sided
device-independent BB84 quantum key distribution with tolerated
bit-error rate $\gamma$ and asymptotically optimal reconciliation,
where $h_2(\cdot)$ is the binary entropy function.  Alice's measurements are trusted,
while Bob may perform an arbitrary joint measurement depending on
the entire basis string, subject to the protocol's receipt and
isolation requirements. This recovers the usual BB84 rate in this
model.
\end{abstract}

\section{Introduction}
\label{sec:introduction}

Monogamy of entanglement limits the information that two separated parties can
simultaneously obtain about the outcomes of incompatible quantum measurements
performed by a third party, Alice.
This restriction connects monogamy-of-entanglement games (monogamy games for short) with quantum key
distribution (QKD). In a monogamy game, two separated players try to 
predict Alice's measurement outcome, but monogamy restricts their simultaneous success.
In QKD, the competing parties are a legitimate receiver and an eavesdropper. When Alice performs trusted incompatible measurements, sufficiently accurate agreement between the legitimate parties constrains the information available to the eavesdropper. A secrecy proof must make this constraint quantitative in the presence
of noise and quantum side information, so that privacy amplification
can turn the raw data into a key that is nearly uniform and independent
of the eavesdropper. Robust monogamy games provide a natural setting
for studying this competition when errors are allowed.

Both problems can be approached through norm bounds on averages of measurement operators.
For fixed measurement strategies, the probability of success in the game
is the expectation of a positive acceptance operator, averaged over
the referee's basis choice. The operator norm of this average bounds
that probability uniformly over the shared state, while pairwise
overlaps of acceptance operators reflect measurement incompatibility.
We develop an inequality that exploits these overlaps by approximating
a related averaging projection. Applied to the game's acceptance projectors,
this inequality gives the sharp robustness threshold for the BB84 monogamy game; applied to a larger class of
quantum tests, it supplies the entropy bound used to establish secrecy in our QKD application. The two
applications thus share a common operator method, with different tests
capturing their respective success and secrecy requirements.

In the following subsections, we introduce the two problems and present our main results, before explaining the operator inequality that underlies both applications.

\subsection{Robust BB84 monogamy game}
\label{sec:intro-robust-monogamy}

The BB84 monogamy-of-entanglement game asks how well two separated
players can simultaneously predict outcomes of incompatible
measurements~\cite{TFKW13}. In this game, two players, Bob and Charlie, prepare a state $\rho_{ABC}$,
give the qubit $A$ to the referee Alice, and retain $B$ and $C$.
Alice chooses $\theta\in\{0,1\}$ uniformly and measures in the basis
$\{\ket{x^\theta}=H^\theta\ket x:x\in\{0,1\}\}$, where $H$ is the Hadamard
matrix. She announces $\theta$, and Bob and Charlie make local
measurements to output bits $y$ and $z$. They win if $x=y=z$.
Their shared state and local measurements are arbitrary, but they cannot
communicate after giving $A$ to Alice.

In the $n$-fold parallel repetition, Alice holds $n$ qubits and chooses
$\theta\in\{0,1\}^n$ uniformly. We write
\begin{equation}
  \ket{x^\theta}:=\bigotimes_{k=1}^n H^{\theta_k}\ket{x_k},
  \qquad x,\theta\in\{0,1\}^n,
  \label{eq:bb84-basis-vector}
\end{equation}
for her measurement basis vectors. Bob and Charlie output strings
$y,z\in\{0,1\}^n$ after learning the entire string $\theta$.
Tomamichel, Fehr, Kaniewski, and Wehner~\cite[Theorem~3]{TFKW13}
proved a strong parallel repetition theorem for this game: the optimal probability that both
players reproduce $x$ exactly, which we denote by $p_{\mathrm{win}}^{(n)}(0,0)$, is given by
\begin{equation}
  p_{\mathrm{win}}^{(n)}(0,0)=\left(\frac12+\frac1{2\sqrt2}\right)^n
  =\left(\cos^2\frac\pi8\right)^n.
  \label{eq:intro-exact-bb84-value}
\end{equation}
This holds even for states entangled across all coordinates and arbitrary
joint measurements within each player's laboratory.

We study the robust version of this game, in which each player may make a bounded number
of errors. Given thresholds $\beta,\gamma\in[0,1/2]$, Bob and Charlie win
when
\begin{equation}
  d_H(x,y)\leq\beta n,
  \qquad d_H(x,z)\leq\gamma n,
  \label{eq:two-sided-winning-condition}
\end{equation}
where $d_H(\cdot, \cdot)$ denotes Hamming distance. We denote the supremum of the
winning probability over all strategies by
$p_{\mathrm{win}}^{(n)}(\beta,\gamma)$.
The imperfect-guessing extension in~\cite[Theorem~4 and Eq.~(9)]{TFKW13}
gives
\begin{equation}
  p_{\mathrm{win}}^{(n)}(\beta,\gamma)
  \leq 2^{n(h_2(\beta)+h_2(\gamma))}
        \left(\frac12+\frac1{2\sqrt2}\right)^n,
  \label{eq:intro-tfkw-robust-bound}
\end{equation}
where $h_2(p)=-p\log p-(1-p)\log(1-p)$ is the binary entropy function, with logarithms to base $2$. This bound proves exponential decay
for sufficiently small error thresholds, but its robustness region is
far from optimal. For example, when $\beta=\gamma$, it certifies
exponential decay only below an error threshold of approximately $1.53\%$.

We completely determine the asymptotic robustness region for this monogamy game in Theorem~\ref{thm:sharp-two-sided-bb84}. We show that if
$(1-2\beta)^2+(1-2\gamma)^2>1$, then
$p_{\mathrm{win}}^{(n)}(\beta,\gamma)$ decays exponentially in $n$. Also, if $(1-2\beta)^2+(1-2\gamma)^2<1$, there is a strategy whose winning
probability tends to $1$ exponentially fast. On the boundary
\begin{equation}
  (1-2\beta)^2+(1-2\gamma)^2=1,
  \label{eq:intro-robustness-boundary}
\end{equation}
there is a strategy with winning probability bounded away from zero.
Thus this circle exactly separates exponential decay from the region
where nonvanishing success is possible. In the symmetric case, the sharp
threshold is
$\beta=\gamma=\sin^2(\pi/8)\approx14.64\%$.

\subsection{One-sided device-independent QKD}
\label{sec:intro-qkd}

Our second application concerns one-sided device-independent QKD.
We study the entanglement-based BB84 protocol E-QKD of~\cite{TFKW13}. Alice prepares
EPR pairs, keeps one half of each pair, and sends the other halves to
Bob. Alice's measurement devices are trusted but the channel from Alice to Bob can be noisy and leak information to Eve. Bob confirms receipt before the basis string is publicly announced by Alice. Bob then applies
an arbitrary joint measurement depending on the entire basis string. Next, a random sample of outcomes is tested
for agreement, and the remaining bits undergo information reconciliation and privacy amplification.

The key ingredient in analyzing this QKD protocol is Proposition~\ref{prop:entropy-rate} establishing the following min-entropy bound. For an arbitrary initial state $\rho_{ABC}$, let $X$ be Alice's BB84 outcome string, $\Theta$ her uniform basis string, and $Y$ Bob's output. On the
subnormalized state $\omega$ restricted to the event $d_H(X,Y)\le\beta n$, with $0\le\beta<1/2$, we prove
that, for every fixed $\varepsilon\in(0,1)$ and $\nu>0$ and all sufficiently
large $n$,
\begin{equation}
 H_{\min}^{\varepsilon}(X|C\Theta)_\omega
 \ge n\bigl(1-h_2(\beta)-\nu\bigr).
 \label{eq:intro-entropy-rate}
\end{equation}
This bound gives a finite-block secrecy guarantee and, with
asymptotically optimal reconciliation, establishes the achievability
of every asymptotic key rate
\begin{equation}
 0<R<1-2h_2(\gamma),
 \label{eq:intro-qkd-rate}
\end{equation}
where $\gamma$ is the tolerated bit-error rate and $R$ is measured in bits per raw qubit; see Theorem~\ref{thm:qkd}. Thus, our bound certifies a positive asymptotic key rate whenever the tolerated bit-error rate $\gamma$ is below approximately $11\%$.

The direct comparison is with the analysis of the same protocol
in~\cite[Theorem~5]{TFKW13}, which certifies rates below
\begin{equation}
 -\log\!\left(\frac12+\frac1{2\sqrt2}\right)-2h_2(\gamma).
 \label{eq:intro-tfkw-key-rate}
\end{equation}
Their zero-noise rate is approximately $0.2284$ bits per raw qubit,
and their noise threshold is approximately $1.53\%$. Our analysis
recovers the usual BB84 rate~\cite{ShorPreskill00} and noise threshold 
but without making any assumptions on Bob's measurements. The rate expression itself also appears in earlier one-sided
device-independent analyses with memoryless measurement devices,
including~\cite{TomamichelRenner11,BranciardEtAl12}. More recently, Bluhm and collaborators~\cite{BluhmEtAl26} developed
numerical methods for one-sided device-independent key rates and
finite-size security bounds based on generalized entropy accumulation.
Their framework allows device memory across rounds within a sequential
protocol. Our analysis instead directly accommodates an arbitrary joint
measurement by Bob, with access to the entire basis string $\Theta$
before any output is required, without imposing a causal ordering
among his individual output bits.

Fully device-independent QKD also leaves Alice's measurements
uncharacterized and therefore uses different certification data and
assumptions. Recent connections include the monogamy-based protocols
of Cervero-Mart\'{i}n and Tomamichel~\cite{CerveroTomamichel25}, the
parallel CHSH protocol of Marwah and Dupuis~\cite{MarwahDupuis25}, and
routed Bell-test protocols that recover BB84 rates in the limit of
ideal local self-tests~\cite{BluhmKossmannSchwonnek26}. These results
provide complementary comparisons concerning measurement trust,
parallel operation, and local certification. Our result concerns
trusted BB84 measurements on Alice's side and arbitrary joint
measurements on Bob's side, under the receipt and isolation
assumptions of E-QKD.

\subsection{The operator bound and proof strategy}
\label{sec:intro-operator-method}

The proof of the strong parallel repetition theorem in~\cite{TFKW13} uses a lemma bounding the operator norm of an average of acceptance projections, $\frac{1}{N}\sum_i A_i$. The first observation is that this operator norm is equal to that of the block matrix $\frac1N(\sqrt{A_i}\sqrt{A_j})_{i,j}$. The proof in~\cite{TFKW13} bounds the pairwise overlaps $\|\sqrt{A_i}\sqrt{A_j}\|$, and combines them through a complete
orthogonal family of permutations to decompose $(\sqrt{A_i}\sqrt{A_j})$ into a sum of $N$ matrices, each of which has at most one nonzero block in each row and each column. 
Our main technical contribution is a
robust extension of this method, stated in
Theorem~\ref{thm:matrix-approximation}. 

For an arbitrary probability distribution $p=(p_1,\ldots,p_N)$, let $R=R_p$ be the rank-one orthogonal projection with entries $R_{ij}=\sqrt{p_i p_j}$. We call $R_p$ the averaging projection associated with $p$. To bound $\opnorm{\sum_i p_i A_i}$, 
we allow an arbitrary approximation $\widetilde{R}$ to $R$ and apply the
permutation decomposition to the corresponding block matrix weighted by $\widetilde{R}$.
The resulting bound has two terms: a spectral approximation error
$\big\|R-\widetilde{R}\big\|\max_i\opnorm{A_i}$ and a sum of pairwise
overlaps $\opnorm{\sqrt{A_i}\sqrt{A_j}}$ weighted by the entries of $\widetilde R$. 
Approximation of $R$ with $\widetilde R$ allows us to 
choose these weights to take nonzero values only when the corresponding overlaps are small. The spectral-error term controls the cost of this approximation. A suitable choice of $\widetilde R$ therefore balances the approximation error against the selected overlaps.

For binary-string indices, we construct $\widetilde{R}$ using a
low-degree polynomial approximation to the NOR function, following the
approach of~\cite{KondraEtAl26}. The resulting matrix is supported on
pairs of strings at Hamming distance close to $n$; see
Theorem~\ref{thm:boolean-cube-bound}. For the robust BB84 monogamy game, the
corresponding projection overlaps reduce to overlaps between
complementary Hamming-ball projections. The uncertainty estimate of
Polyanskiy and Samorodnitsky~\cite{PolyanskiySamorodnitsky19} makes these
overlaps exponentially small throughout the region
$(1-2\beta)^2+(1-2\gamma)^2>1$. Combining these ingredients proves the
sharp converse bound for the monogamy game.

For the min-entropy bound~\eqref{eq:intro-entropy-rate}, we replace
the third player's Hamming-ball test by arbitrary effects whose sum
is bounded by $L$ times the identity operator. These effects come from a relaxation of the problem of 
quantum list decoding with list size $L$.
An overlap estimate for this class of operators, combined with the same
Boolean-cube bound, shows that its optimal acceptance value tends
to zero as $n\to\infty$ whenever
$\log L<n(1-h_2(\beta)-\nu)$ for a fixed $\nu>0$.
Semidefinite duality then converts this bound into~\eqref{eq:intro-entropy-rate}.

Our operator bound via matrix approximation has other applications. We also apply this bound to the subspace
coset monogamy problem introduced by Coladangelo, Liu, Liu, and
Zhandry~\cite{ColadangeloEtAl21}. They conjectured its
information-theoretic monogamy property, which Culf and
Vidick~\cite{CulfVidick22} subsequently proved by establishing an
exponentially small upper bound on the winning probability.
Here, instead of bit strings, the questions are $n$-dimensional subspaces of a $2n$-dimensional vector space over $\mathbb F_2$. In this case, we approximate the averaging matrix $R$
by a matrix $\widetilde{R}$ supported on complementary subspaces that have zero intersection. 
With this choice of $\widetilde{R}$, Corollary~\ref{cor:coset-winning-bound} gives
\begin{align}
  p_{\mathrm{coset}}^{(2n)}\leq 2^{-n}+2^{-\frac n2}.
\end{align}
This recovers the same exponent as established by Schleppy and
Soljanin~\cite{SchleppySoljanin25}, but with an improved prefactor. 
We refer to Appendix~\ref{sec:grassmannian-coset} for a precise formulation of the subspace
coset monogamy game and more details about our proof method.

\subsection{Organization of the paper}
\label{sec:intro-organization}

We begin in Section~\ref{sec:shaped-operator-norm-bound} by proving
our general operator-norm inequality and its specialization to
the Boolean cube. We then turn to its two main applications.
In Section~\ref{sec:robust-bb84}, we establish the sharp robustness
threshold for the BB84 monogamy game.
Sections~\ref{sec:entropy-proof} and~\ref{sec:qkd-proof} develop
the QKD application: the former derives a smooth min-entropy
bound through a relaxation of quantum list decoding, and the
latter uses this bound to prove secrecy and obtain the achievable
key rates. 
Appendix~\ref{app:elementary-ball-overlap} provides a direct proof
of the complementary Hamming-ball overlap estimate used in the
monogamy analysis, while Appendix~\ref{sec:grassmannian-coset}
presents a further application of our operator inequality
to subspace coset monogamy games.

\section{Operator bounds via approximation of the averaging projection}
\label{sec:shaped-operator-norm-bound}

This section develops the main operator-norm estimate and then specializes it to operators indexed by the Boolean cube.  This specialization is the form used in Sections~\ref{sec:robust-bb84} and~\ref{sec:entropy-proof}.

All logarithms are to base $2$, except $\ln$, which denotes the natural logarithm. We use the binary entropy $h_2(x) = -x \log x - (1-x)\log (1-x)$, with $0\log 0=0$. For nonnegative integers $k$, we set $\binom{k}{j}=0$ when $j<0$ or $j>k$, and interpret empty sums as zero. We write $[n]:=\{1,\ldots,n\}$, and for $u,v\in\{0,1\}^n$, we use $d_H(u,v)$ for their Hamming distance.
Also, the Hamming weight of $u\in \{0,1\}^n$ is $|u|:=d_H(0^n,u)=\sum_{i=1}^n u_i$. For a string $x\in \{0,1\}^n$ and a subset $J\subseteq [n]$ we denote the substring consisting of indices in $J$ by $x_J$.

 All Hilbert spaces considered in this paper are assumed to be finite-dimensional.  For a Hilbert space $\mathcal H$, we denote its identity operator by $I_{\mathcal H}$. Subscripts are omitted when they are clear from context. Throughout, $\opnorm{\cdot}$ denotes the operator norm.
 
We denote the qubit Pauli operators by $X,Y,Z$ and the Hadamard matrix by $H$. The vectors $\ket{\pm}=(\ket0\pm\ket1)/\sqrt2$ are eigenvectors of $X$ with eigenvalues $\pm1$, respectively. 

For two Hermitian operators $A, B$ we write $A\geq B$ if $A-B$ is positive semidefinite. We frequently use the following simple observation: if Hermitian projections $P, P', Q, Q'$ satisfy $P\leq P'$ and $Q\leq Q'$, then we have $\opnorm{PQ}\leq \opnorm{P'Q'}$. This is because $PQ = PP'Q'Q$, so the submultiplicativity of operator norm and $\|P\|, \|Q\|\leq 1$ imply 
\begin{align}\label{eq:intro-PQ-bound}
\opnorm{PQ} = \opnorm{PP'Q'Q}\leq \opnorm{P'Q'}.
\end{align}

\subsection{A general bound for operator averages}

Two permutations $\pi,\sigma$ of $[N]$
are called \emph{orthogonal} if
\begin{equation}
  \label{eq:orthogonal-permutations}
  \pi(i)\neq \sigma(i)
  \qquad\text{for every }i\in[N].
\end{equation}
A family $\mathcal P=\{\pi_1,\ldots,\pi_N\}$ of $N$ mutually orthogonal
permutations will be called a \emph{complete orthogonal family}. For every
$(i,j)\in[N]^2$, there is then a unique $\pi\in \mathcal P$ such that
$\pi(i)=j$.  Such families always exist: after identifying $[N]$ with
$\mathbb Z_N$, the cyclic translations $\pi_k(i)=i+k$ modulo $N$ form a
complete orthogonal family.

\begin{theorem}[Operator-norm bound via matrix approximation]\label{thm:matrix-approximation}
  Let $A_1,\ldots,A_N$ be positive semidefinite operators on
  a finite-dimensional Hilbert space $\cH$, and let $p_1,\ldots,p_N\geq0$
  satisfy $\sum_{i=1}^N p_i=1$. Define $R=R_p\in\mathbb{C}^{N\times N}$
  by $R_{ij}=\sqrt{p_i p_j}$, let $\widetilde{R}=(\tilde{r}_{i,j})\in\mathbb{C}^{N\times N}$ be
  arbitrary, and let $\mathcal P$ be a complete orthogonal family of permutations
  of $[N]$. Then
  \begin{equation}
    \label{eq:main-bound}
    \opnorm{\,\sum_{i=1}^N p_i A_i\,}
    \leq
    \opnorm{R-\widetilde{R}}\max_{i\in[N]}\opnorm{A_i}
    +
    \sum_{\pi\in \mathcal P}
    \max_{i\in[N]}
    \left\{
      |\tilde{r}_{i,\pi(i)}|
      \opnorm{\sqrt{A_i}\sqrt{A_{\pi(i)}}}
    \right\}.
  \end{equation}
\end{theorem}

This generalizes~\cite[Lemma 2]{TFKW13}, which can be recovered by setting~$\widetilde{R} = R$.
Theorem~\ref{thm:matrix-approximation} is also reminiscent of, and motivated
by~\cite[Lemma~S1]{KondraEtAl26}, but gives a more
general formulation and a stronger overlap dependence in the setting
used there with uniform weights. Roughly speaking,~\cite[Lemma~S1]{KondraEtAl26} establishes a similar inequality to~\eqref{eq:main-bound} but with the second term replaced by its square root. Thus~\cite[Lemma~S1]{KondraEtAl26} is weaker than Theorem~\ref{thm:matrix-approximation} when the second term is smaller than $1$, which is often the case in applications.
Our theorem also applies to arbitrary finite
index sets and probability distributions, positive semidefinite operators, and complex approximation
matrices, without the symmetry and row-sum normalization assumptions imposed in~\cite[Lemma~S1]{KondraEtAl26}.

\begin{proof}
  Define the block-column operator $\Lambda:\cH\to\cH^{\oplus N}$ and the
  block-diagonal operator $\Delta$ on $\cH^{\oplus N}$ by
  \begin{equation}
    \label{eq:def-X}
    \Lambda=
    \begin{pmatrix}
      \sqrt{p_1}\sqrt{A_1}\\
      \vdots\\
      \sqrt{p_N}\sqrt{A_N}
    \end{pmatrix},
    \qquad
    \Delta
    :=
    \bigoplus_{i=1}^N\sqrt{A_i}.
  \end{equation}
  Since $\Lambda^\dagger\Lambda=\sum_{i=1}^N p_i A_i$, direct block multiplication gives
  \begin{equation}
    \big(\Lambda\Lambda^\dagger\big)_{ij}=\sqrt{p_i p_j}\sqrt{A_i}\sqrt{A_j},
    \qquad
    \Lambda\Lambda^\dagger=\Delta(R\otimes I_{\cH})\Delta.
    \label{eq:x-x-star-block}
  \end{equation}
  Consequently,
  \begin{equation}
    \label{eq:block-trick}
    \opnorm{\,\sum_{i=1}^N p_i A_i\,}
    =
    \opnorm{\Lambda^\dagger\Lambda}
    =
    \opnorm{\Lambda\Lambda^\dagger}
    =
    \opnorm{\Delta(R\otimes I_{\cH})\Delta}.
  \end{equation}
Now adding and subtracting
  $\Delta(\widetilde{R}\otimes I_{\cH})\Delta$ and applying the triangle inequality gives
  \begin{equation}
    \label{eq:triangle}
    \opnorm{\Delta(R\otimes I_{\cH})\Delta}
    \leq
    \opnorm{\Delta((R-\widetilde{R})\otimes I_{\cH})\Delta}
    +
    \opnorm{\Delta(\widetilde{R}\otimes I_{\cH})\Delta}.
  \end{equation}
  Since
  $\opnorm{\Delta}^2=\max_i\opnorm{A_i}$, the first term satisfies
  \begin{equation}
    \label{eq:error-term}
    \opnorm{\Delta((R-\widetilde{R})\otimes I_{\cH})\Delta}
    \leq
    \opnorm{R-\widetilde{R}}\max_{i\in[N]}\opnorm{A_i}.
  \end{equation}

  For each $\pi\in \mathcal P$, let 
  \begin{equation}
    \label{eq:def-d-pi}
    D_\pi
    :=  \sum_{i=1}^N \tilde{r}_{i,\pi(i)} \ketbra{i}{\pi(i)} \otimes \sqrt{A_i}\sqrt{A_{\pi(i)}}.
  \end{equation}
  Completeness and mutual orthogonality of
  the permutations imply
  $\Delta(\widetilde{R}\otimes I_{\cH})\Delta=\sum_{\pi\in \mathcal P}D_\pi$, and hence
  \begin{align}
    \opnorm{\Delta(\widetilde{R}\otimes I_{\cH})\Delta}
    \leq
    \sum_{\pi\in \mathcal P}\opnorm{D_\pi}.
    \label{eq:shaped-term-triangle}
   \end{align}
 We also note that each row and column of $D_\pi$ has at most one nonzero block. Therefore,  
   \begin{align}
    \sum_{\pi\in \mathcal P}\opnorm{D_\pi}
    =
    \sum_{\pi\in \mathcal P}
    \max_{i\in[N]}
    \left\{
      |\tilde{r}_{i,\pi(i)}|
      \opnorm{\sqrt{A_i}\sqrt{A_{\pi(i)}}}
    \right\}.
    \label{eq:shaped-term-bound}
  \end{align}
  Combining \eqref{eq:block-trick}, \eqref{eq:triangle},
  \eqref{eq:error-term}, \eqref{eq:shaped-term-triangle}, and
  \eqref{eq:shaped-term-bound} proves \eqref{eq:main-bound}.

\end{proof}

\medskip

Taking $\widetilde{R}=R$ makes the first term on the right-hand side of
  \eqref{eq:main-bound} vanish and gives
  \begin{equation}
    \label{eq:tfkw-bound}
    \opnorm{\sum_{i=1}^N p_i A_i}
    \leq
    \sum_{\pi\in \mathcal P}
    \max_{i\in[N]}
    \left\{\sqrt{p_i p_{\pi(i)}}
    \opnorm{\sqrt{A_i}\sqrt{A_{\pi(i)}}}\right\}.
  \end{equation}
  For $p_i=1/N$, this is the averaged form of~\cite[Lemma~2]{TFKW13}.

\subsection{Operator averages indexed by the Boolean cube}
\label{sec:boolean-shaping-matrix}

In applications, our operators are often indexed by binary strings $\theta\in \{0,1\}^n$ corresponding to measurement bases, and the norms $\opnorm{\sqrt{A_\theta}\sqrt{A_{\theta'}}}$ are bounded by a
function of the Hamming distance $d_H(\theta,\theta')=|\theta\oplus\theta'|$.
The following theorem combines Theorem~\ref{thm:matrix-approximation} with
a polynomial approximation on the Boolean cube to obtain a bound on the average of $A_\theta$'s under these conditions.

\begin{theorem}[Boolean-cube operator-norm bound]\label{thm:boolean-cube-bound}
Let $A_\theta$, for $\theta\in\{0,1\}^n$, be positive semidefinite operators on a common Hilbert space, and suppose that
$g:\{0,\ldots,n\}\to[0,\infty)$ satisfies
\begin{equation}\label{eq:boolean-cube-overlap-profile}
 \opnorm{\sqrt{A_\theta}\sqrt{A_{\theta'}}}
 \le g(d_H(\theta,\theta'))\qquad\text{for all }\theta,\theta'.
\end{equation}
There is a universal constant $c>0$ such that, for every integer $3\le D\le n$,
\begin{equation}\label{eq:boolean-cube-final-bound}
 \opnorm{2^{-n}\sum_\theta A_\theta}
 \le 2^{-cD^2/n}\max_\theta\opnorm{A_\theta}
 +\left(\sum_{k=0}^D\binom nk\right)^{1/2}
\left( \max_{n-D\le s\le n}g(s)\right).
\end{equation}
In particular, if $g(k)\le\kappa2^{-\alpha k}$ for $n-D\le k\le n$,
where $\kappa,\alpha>0$, then
\begin{equation}\label{eq:boolean-cube-exponential-bound}
 \opnorm{2^{-n}\sum_\theta A_\theta}
 \le 2^{-cD^2/n}\max_\theta\opnorm{A_\theta}
 +\kappa\left(\sum_{k=0}^D\binom nk\right)^{1/2}2^{-\alpha(n-D)}.
\end{equation}
\end{theorem}

To prove this theorem we use Theorem~\ref{thm:matrix-approximation}. We construct the approximating matrix by applying a low-degree polynomial to a sum of commuting local projections. As in~\cite{KondraEtAl26}, the
polynomial ingredient is the following approximation of NOR. 

\begin{lemma}[NOR approximation {\cite[Theorem~4.5]{Sherstov20}}]
 \label{lem:nor-polynomial}
There is a universal constant $c>0$ such that, for integers $3\le D\le n$,
there is a real polynomial $q_{n,D}$ of degree at most $D$ satisfying
\begin{equation}
 \label{eq:nor-approximation}
 q_{n,D}(0)=1,\qquad
 \max_{k\in\{1,\ldots,n\}}|q_{n,D}(k)|\le2^{-cD^2/n},\qquad
 \max_{k\in\{0,\ldots,n\}}|q_{n,D}(k)|\le1.
\end{equation}
\end{lemma}

\begin{proof}[Proof of Theorem~\ref{thm:boolean-cube-bound}]
We apply Theorem~\ref{thm:matrix-approximation} with the uniform distribution
$p_\theta=2^{-n}$. In this case the corresponding averaging projection is a tensor product:
\begin{equation}\label{eq:boolean-product-projection}
 R=\bigl(\ketbra{+}{+}\bigr)^{\otimes n},
 \qquad \ket{\pm}=\frac{\ket0\pm\ket1}{\sqrt2}.
\end{equation}
Then following the approach of~\cite{BeigiTomamichel26} and references therein, we can approximate the tensor product projection $R$ using the above lemma.

Let $X=\ketbra01+\ketbra10$ be the Pauli bit flip, and let $X_i$ act as $X$
on the $i$th tensor factor and as the identity elsewhere. Thus, $\frac{I\pm X_i}{2} = \ketbra{\pm}{\pm}_{i}$. Define
\begin{equation}\label{eq:boolean-defect-operator}
 L:=\sum_{i=1}^n \frac{I- X_i}{2} = \sum_{i=1}^n \ketbra{-}{-}_i,
 \end{equation}
 and note that 
 \begin{align}
  R=\prod_{i=1}^n \ketbra{+}{+}_i = \prod_{i=1}^n \frac{I+X_i}{2}.
 \end{align}
The projections $\frac{I- X_i}{2}$ commute. Moreover, on a joint eigenvector of these projections, $L$ counts the
number of factors on which $\frac{I- X_i}{2}$ has eigenvalue $1$. Thus the spectrum of $L$ is
$\{0,1,\ldots,n\}$. Also, $R$ is precisely its zero-eigenspace projection.

For the polynomial $q_{n,D}$ from Lemma~\ref{lem:nor-polynomial}, set 
\begin{align}
\widetilde{R} = X^{\otimes n} q_{n, D}(L).
\end{align}
Since $X^{\otimes n}R=R$ and $X^{\otimes n}$ is unitary, by examining the spectrum of $L$ we find that
\begin{align}\label{eq:boolean-polynomial-error}
 \opnorm{\widetilde{R}-R\,}
 &=\opnorm{q_{n,D}(L)-R}
 =\max_{1\le k\le n}|q_{n,D}(k)|\le2^{-cD^2/n},\\
 \opnorm{\widetilde{R}\,}&=\opnorm{q_{n,D}(L)}=\max_{0\leq k\leq n} |q_{n,D}(k)| = q_{n, D}(0)=1.
\end{align}

Because $q_{n,D}$ has degree at most $D$, expanding
$q_{n,D}(L)=q_{n,D}(\sum_i(I-X_i)/2)$, we obtain terms that are products of at most $D$ of the operators $X_i$.
Using $X_i^2=I$ and their commutativity, we can therefore write
\begin{equation}\label{eq:boolean-local-expansion}
 q_{n, D}(L)=\sum_{\substack{J\subseteq[n]\\|J|\le D}}a_J X_J,
 \qquad X_J:=\prod_{i\in J}X_i,
\end{equation}
with real coefficients $a_J$ and $X_\varnothing=I$. Hence, applying $q_{n,D}(L)$ to any computational-basis state $\ket \theta$, we obtain a linear combination of vectors $\ket {\theta'}$ satisfying $d_H(\theta, \theta')\leq D$. Next, applying $X^{\otimes n}$ flips all the bits, so the entries of $\widetilde{R}=X^{\otimes n}q_{n,D}(L)$ satisfy 
\begin{equation}\label{eq:shaping-support}
 \tilde r_{\theta,\theta'}=0
 \qquad\text{ if } \qquad d_H(\theta,\theta')<n-D.
\end{equation}

Both $q_{n, D}(L)$ and $X^{\otimes n}$ are invariant under permutations of the tensor
factors and commute with every bit flip $X_J$. Consequently, the entries of
$\widetilde{R}$ depend only on Hamming distance. In other words,
$\tilde r_{\theta,\theta'}$ depends only on $\vartheta=\theta\oplus\theta'$, so we write $\tilde r_\vartheta:=\tilde r_{0^n,\vartheta}$. Then,~\eqref{eq:shaping-support} implies that $\tilde r_{\vartheta}=0$ if $|\vartheta|<n-D$.  

In order to apply Theorem~\ref{thm:matrix-approximation}, take the complete orthogonal family of permutations 
$\mathcal P=\{\pi_\vartheta:\vartheta\in\{0,1\}^n\}$, where
$\pi_\vartheta(\theta)=\theta\oplus\vartheta$.
For each permutation $\pi_\vartheta$, we have
$\tilde r_{\theta,\pi_\vartheta(\theta)}=\tilde r_\vartheta$ and
$d_H(\theta,\pi_\vartheta(\theta))=|\vartheta|$. Thus the second term on the
right-hand side of~\eqref{eq:main-bound} satisfies
\begin{align}
 \sum_\vartheta\max_\theta
 \left\{|\tilde r_{\theta,\pi_\vartheta(\theta)}|
 \opnorm{\sqrt{A_\theta}\sqrt{A_{\pi_\vartheta(\theta)}}}\right\}& =  \sum_\vartheta |\tilde r_{\vartheta}| \max_\theta
 \opnorm{\sqrt{A_\theta}\sqrt{A_{\theta\oplus\vartheta}}} \\
 &\leq\sum_\vartheta|\tilde r_\vartheta|g(|\vartheta|)\\
  &=\sum_{\vartheta:\,|\vartheta|\ge n-D}|\tilde r_\vartheta|g(|\vartheta|)\\
 &\leq\left(\max_{n-D\le s\le n}g(s)\right)
       \sum_{\vartheta:\,|\vartheta|\ge n-D}|\tilde r_\vartheta|.
\end{align}
We note that there are at most $\sum_{k=0}^D\binom nk$ nonzero values of $\tilde r_{\vartheta}$. Thus, applying
the Cauchy--Schwarz inequality, we have
\begin{align}
 \sum_{\vartheta: |\vartheta|\geq n-D} | \tilde r_{\vartheta}| &\le\left(\,\sum_{k=0}^D\binom nk\,\right)^{1/2}
       \left(\sum_{\vartheta}|\tilde r_{\vartheta}|^2\right)^{1/2}\\
 &=\left(\,\sum_{k=0}^D\binom nk\,\right)^{1/2}
       \big\|\widetilde{R}\ket{0^n}\big\|_2
 \le\left(\sum_{k=0}^D\binom nk\right)^{1/2},
 \label{eq:shaping-row-l1}
\end{align}
where the last step uses $\big\|\widetilde{R}\big\|\le1$. Applying this bound in Theorem~\ref{thm:matrix-approximation} and using~\eqref{eq:boolean-polynomial-error} gives~\eqref{eq:boolean-cube-final-bound}. The exponential specialization
follows from $2^{-\alpha s}\le2^{-\alpha(n-D)}$ for $n-D\le s\le n$.

\end{proof}

\section{Sharp robustness thresholds for the BB84 monogamy game}
\label{sec:robust-bb84}

We now apply Theorem~\ref{thm:boolean-cube-bound} to determine the sharp robustness
threshold for the BB84 monogamy game introduced in
Section~\ref{sec:introduction}. The converse reduces overlaps of the
game's acceptance projectors to overlaps between complementary
Hamming-ball projectors, which are controlled by an uncertainty estimate.
To show sharpness, we construct a strategy based on an asymmetric qubit
cloner and analyze its independent repetitions, including the boundary
of the robustness region. Throughout, we use the winning condition
\eqref{eq:two-sided-winning-condition} and the notation
$p_{\mathrm{win}}^{(n)}(\beta,\gamma)$ established in the introduction.

\begin{theorem}[Sharp two-sided robustness threshold]
  \label{thm:sharp-two-sided-bb84}
  Let $\beta,\gamma\in [0, 1/2]$ be the error thresholds for Bob and Charlie, respectively. If
  \begin{equation}
      \left(\frac12-\beta\right)^2 + \left(\frac12-\gamma\right)^2>\frac 14,
    \label{eq:two-sided-outside-circle}
  \end{equation}
  then there is a constant $c_{\beta,\gamma}>0$ such that
  \begin{equation}
    p_{\mathrm{win}}^{(n)} (\beta,\gamma) \leq 2^{-c_{\beta,\gamma}n}
    \label{eq:two-sided-exponential-decay}
  \end{equation}
  for all sufficiently large $n$.  If the reverse strict inequality holds,
  there is a strategy with winning probability tending to $1$ exponentially
  fast.  On the boundary
  \begin{equation}
  \left(\frac12-\beta\right)^2
  +
  \left(\frac12-\gamma\right)^2
  =
  \frac14,
  \label{eq:two-sided-circle-boundary}
\end{equation}
 there is a strategy with winning probability bounded away from zero.
\end{theorem}

It is instructive to examine some special cases of this theorem. In the case of symmetric thresholds, the boundary values are $\beta=\gamma  = \sin^2\!\left(\frac{\pi}{8}\right)$, which correspond to the strategy of always outputting $0$ while preparing Alice's state in $\cos\!\left(\frac{\pi}{8}\right)\ket 0 + \sin\!\left(\frac{\pi}{8}\right)\ket 1$.  In this case, Alice's outcome is equal to $0$ with probability $\cos^2\!\left(\frac{\pi}{8}\right)$ independent of the measurement basis $\theta$, giving an error probability of $1-\cos^2\!\left(\frac{\pi}{8}\right)= \sin^2\!\left(\frac{\pi}{8}\right)$.

In the extreme case of $\beta=0$ and $\gamma=1/2$, Bob can share an EPR pair with Alice in which case he can output Alice's string perfectly. Charlie can output a uniformly random string, guessing half of Alice's bits correctly on average.

The proof of Theorem~\ref{thm:sharp-two-sided-bb84} is based on an uncertainty result due to Polyanskiy and Samorodnitsky \cite[Theorem~9]{PolyanskiySamorodnitsky19}. On $n$ qubits, for any real radius $k\geq0$, define the two
complementary Hamming-ball projectors
\begin{equation}
  \Pi^{(n)}_{k}:=\sum_{|x|\leq k}\lvert x\rangle\!\langle x\rvert
  \label{eq:z-ball-projector}
\end{equation}
and
\begin{equation}
  \widehat\Pi^{(n)}_{k}:= H^{\otimes n}\Pi^{(n)}_{k}(H^\dagger)^{\otimes n}.
  \label{eq:x-ball-projector}
\end{equation}
Here $H$ is the Hadamard matrix and $\widehat\Pi^{(n)}_k$ is the Fourier conjugate of $\Pi^{(n)}_k$. For $k\geq n$
both $\Pi^{(n)}_{k}$ and $\widehat\Pi^{(n)}_{k}$ are the identity operator, and for $n=0$ both projections are the scalar $1$.

\begin{lemma} \textup{\cite[Theorem~9]{PolyanskiySamorodnitsky19}}
  \label{lem:Polyanskiy-Samorodnitsky}
  For every $\beta,\gamma\in[0,1/2]$ satisfying
\begin{equation}
  (1-2\beta)^2+(1-2\gamma)^2>1,
  \label{eq:ps-uncertainty-condition}
\end{equation}
there is $\varepsilon=\varepsilon_{\beta,\gamma}>0$ such that for all sufficiently large $n$ we have
\begin{equation}
  \opnorm{\Pi^{(n)}_{\mathcal{S}_\gamma}\widehat\Pi^{(n)}_{\beta n}}\leq 2^{-\varepsilon n},
  \label{eq:ps-hamming-ball-overlap}
\end{equation}
where $\mathcal{S}_{\gamma}$ is an arbitrary subset of $\{0,1\}^n$ satisfying $|\mathcal{S}_{\gamma}| \leq |\{x\in \{0,1\}^n\:\, |x|\leq \gamma n\}|$ and $\Pi^{(n)}_{\mathcal{S}_\gamma} = \sum_{x\in \mathcal{S}_\gamma} \ketbra xx$.
\end{lemma}

We only need a special case of this lemma in which $\mathcal{S}_{\gamma}=\{x\in \{0,1\}^n\:\, |x|\leq \gamma n\}$ and $\Pi^{(n)}_{\mathcal{S}_\gamma} = \Pi^{(n)}_{\gamma n}$. In Appendix~\ref{app:elementary-ball-overlap}
we give a short direct proof of this special case with an explicit exponential decay rate.

Fix an arbitrary strategy.  After local Naimark dilations, write Bob and Charlie’s
measurements as projective measurements $\big\{B_y^\theta:\, y\in \{0,1\}^n\big\}$ and
$\big\{C_z^\theta:\, z\in \{0,1\}^n\big\}$. Using the basis vectors in~\eqref{eq:bb84-basis-vector},
the acceptance projection for a basis string $\theta\in\{0,1\}^n$ is
\begin{equation}
  A_\theta:=\sum_{x\in\{0,1\}^n}  \lvert x^\theta\rangle\!\langle x^\theta\rvert
  \otimes \left(\sum_{d_H(x,y)\leq \beta n}B_y^\theta \right)
  \otimes \left( \sum_{d_H(x,z)\leq \gamma n}C_z^\theta \right).
  \label{eq:two-sided-acceptance-projection}
\end{equation}
Then, letting $\rho$ be the shared state in the strategy, its winning probability satisfies
\begin{equation}
  p_{\mathrm{win}}= \frac 1{2^{n}}\sum_\theta\operatorname{Tr}[\rho A_\theta]
  \leq
  \opnorm{\frac 1{2^n}\sum_\theta A_\theta}.
  \label{eq:two-sided-winning-probability-norm}
\end{equation}
To bound the left-hand side, we need to estimate the pairwise overlaps of the projections in order to apply Theorem~\ref{thm:boolean-cube-bound}.

\begin{lemma}
  \label{lem:two-sided-overlap-reduction}
  If $d=d_H(\theta,\theta')$, then
  \begin{equation}
    \opnorm{\sqrt{A_\theta}\sqrt{A_{\theta'}}}
    \leq
    \opnorm{\Pi^{(d)}_{\gamma n}\widehat\Pi^{(d)}_{\beta n}}.
    \label{eq:two-sided-acceptance-overlap-reduction}
  \end{equation}
\end{lemma}

\begin{proof}
If $d=0$, the right-hand side is $1$ and the claim trivially holds, so we assume $d\ge1$ in the sequel.
Drop Bob's acceptance condition from $A_\theta$ and define
\begin{align}
  P&=\sum_{x\in\{0,1\}^n}  \lvert x^\theta\rangle\!\langle x^\theta\rvert
  \otimes I_B
  \otimes \left( \sum_{d_H(x,z)\leq \gamma n}C_z^\theta \right)\\
  &  = \sum_{z\in \{0,1\}^n}\underbrace{\left(\sum_{x:\,d_H(x,z)\le \gamma n}|x^\theta\rangle\!\langle x^\theta|\right)}_{=:P_z}\otimes I_B\otimes C_z^\theta.
\end{align}
Similarly drop Charlie's acceptance
condition from $A_{\theta'}$ and define
\begin{align}
 Q=\sum_yQ_y\otimes B_y^{\theta'}\otimes I_C, \qquad Q_y&:=\sum_{x':\,d_H(x',y)\le \beta n}|x'^{\theta'}\rangle\!\langle x'^{\theta'}|.
\end{align}
These projections satisfy $A_\theta\le P$ and $A_{\theta'}\le Q$. Consequently, by our observation in~\eqref{eq:intro-PQ-bound} we have 
\begin{equation}\label{eq:two-sided-overlap-enlargement}
 \opnorm{\sqrt{A_\theta}\sqrt{A_{\theta'}}}
 =\opnorm{A_\theta A_{\theta'}}\le\opnorm{PQ}.
\end{equation}
Because Bob's and Charlie's measurements are projective and $B_y^{\theta'}\otimes C_z^\theta$ for different values of $(y, z)$ are mutually
orthogonal, we have
\begin{equation}\label{eq:outcome-block-decomposition}
  \opnorm{PQ}=\opnorm{\sum_{y,z}P_zQ_y\otimes B_y^{\theta'}\otimes C_z^\theta}\leq\max_{y,z}\opnorm{P_zQ_y}.
\end{equation}

Put $J=\{k:\theta_k\neq\theta'_k\}$, so $|J|=d$. If $x$ and $x'$ differ on any coordinate in $J^c$, then $\braket{x^\theta}{x'^{\theta'}}=0$. Regrouping the tensor factors into $J^c$ and $J$, we can therefore write
\begin{align}
P_zQ_y = \sum_{w_{J^c}\in \{0,1\}^{n-d}} \ketbra{w_{J^c}^{\theta_{J^c}}}{w_{J^c}^{\theta_{J^c}}}\otimes \left(\sum_{x_J\in \mathcal{S}_{w_{J^c}, z}} \ketbra{x_J^{\theta_J}}{x_J^{\theta_J}} \right) \left(\sum_{x'_J\in \mathcal{S}'_{w_{J^c}, y}} \ketbra{{x'_J}^{\theta'_J}}{{x'_J}^{\theta'_J}} \right)
\end{align}
where 
\begin{align}
&\mathcal{S}_{w_{J^c}, z} = \big\{x_J\in \{0,1\}^d:\, d_H(x_J, z_J) + d_H(w_{J^c}, z_{J^c})\leq \gamma n\big\},\\
&\mathcal{S}'_{w_{J^c}, y} = \big\{x'_J\in \{0,1\}^d:\, d_H(x'_J, y_J) + d_H(w_{J^c}, y_{J^c})\leq \beta n\big\}.
\end{align}
Thus, relaxing the conditions defining the sets $\mathcal{S}_{w_{J^c}, z}$ and $\mathcal{S}'_{w_{J^c}, y}$, and once again using~\eqref{eq:intro-PQ-bound} yields
\begin{align}
\opnorm{P_zQ_y } &= \max_{w_{J^c}} \opnorm{\left(\sum_{x_J\in \mathcal{S}_{w_{J^c}, z}} \ketbra{x_J^{\theta_J}}{x_J^{\theta_J}} \right) \left(\sum_{x'_J\in \mathcal{S}'_{w_{J^c}, y}} \ketbra{{x'_J}^{\theta'_J}}{{x'_J}^{\theta'_J}} \right)}\\
& \leq  \opnorm{\left(\sum_{x_J: d_H(x_J, z_J)\leq \gamma n} \ketbra{x_J^{\theta_J}}{x_J^{\theta_J}} \right) \left(\sum_{x'_J: d_H(x'_J, y_J)\leq \beta n} \ketbra{{x'_J}^{\theta'_J}}{{x'_J}^{\theta'_J}} \right)}.\label{eq:relaxed-Pz-Qy}
\end{align}
Conjugating the operator in~\eqref{eq:relaxed-Pz-Qy} by $\bigotimes_{k\in J}H^{\theta_k}$, we can assume with no loss of generality that $\theta_J=0^d$ and $\theta'_J = 1^d$. Next, conjugation by the Pauli operator $X^{z_J}Z^{y_J}:=\bigotimes_{k\in J}X^{z_k}Z^{y_k}$ shows that we can further assume that $z_J=y_J =0^d$. Therefore, 
\begin{align}
 \opnorm{\sqrt{A_\theta}\sqrt{A_{\theta'}}}\leq \max_{y, z}\opnorm{P_zQ_y }  \leq \opnorm{\Pi^{(d)}_{\gamma n}\widehat\Pi^{(d)}_{\beta n}},
\end{align}
as desired.

\end{proof}

\begin{proof}[Proof of Theorem~\ref{thm:sharp-two-sided-bb84}]
  We first prove the upper bound. Because
  \eqref{eq:two-sided-outside-circle} is strict, choose
  $\overline\beta>\beta$ and
  $\overline\gamma>\gamma$, still at most $1/2$, such that
  \begin{equation}
    (1-2\overline\beta)^2
    +(1-2\overline\gamma)^2
    >1.
    \label{eq:enlarged-two-sided-rates}
  \end{equation}
  Let $\varepsilon=\varepsilon_{\overline\beta,\overline\gamma}>0$ be supplied by
  \eqref{eq:ps-hamming-ball-overlap} for this pair.  Choose $0<\delta<1/2$ so
  small that
  \begin{equation}
    \frac{\beta}{1-\delta}<\overline\beta,
    \qquad
    \frac{\gamma}{1-\delta}<\overline\gamma,
    \qquad
    \varepsilon(1-\delta)>\frac12h_2(\delta).
    \label{eq:two-sided-delta-conditions}
  \end{equation}
  Set $D=\lfloor\delta n\rfloor$.  For every $d\geq n-D\geq (1-\delta)n$ and all sufficiently large $n$, Lemma~\ref{lem:Polyanskiy-Samorodnitsky} gives
  \begin{equation}
    \opnorm{\Pi^{(d)}_{\gamma n}\widehat\Pi^{(d)}_{\beta n}}\leq \opnorm{\Pi^{(d)}_{\overline \gamma d}\widehat\Pi^{(d)}_{\overline \beta d}}
    \leq
    2^{-\varepsilon d}
    \leq
    2^{-\varepsilon(n-D)}.
    \label{eq:uniform-two-sided-large-sector-overlap}
  \end{equation}
Then, by Lemma~\ref{lem:two-sided-overlap-reduction} we have
\begin{equation}
    \opnorm{\sqrt{A_\theta}\sqrt{A_{\theta'}}}
    \leq 2^{-\varepsilon(n-D)} \qquad  \text{ if } \qquad d_H(\theta, \theta')\geq n-D.
  \end{equation}
 Hence, applying Theorem~\ref{thm:boolean-cube-bound} yields
   \begin{equation}
    p_{\mathrm{win}} \leq
  \opnorm{\frac 1{2^n}\sum_\theta A_\theta}
    \leq
    2^{-cD^2/n}
    +
    \left(\sum_{j=0}^D\binom nj\right)^{1/2}
    2^{-\varepsilon(n-D)}.
    \label{eq:two-sided-finite-block-bound}
  \end{equation}
  The standard Hamming-ball size estimate bounds the second term by
  \begin{equation}
    2^{-n\left(
      \varepsilon(1-\delta)-\frac12h_2(\delta)
    \right)+o(n)}.
    \label{eq:two-sided-overlap-exponent}
  \end{equation}
  Thus, both terms in \eqref{eq:two-sided-finite-block-bound} decay exponentially by
  \eqref{eq:two-sided-delta-conditions}, proving
  \eqref{eq:two-sided-exponential-decay} uniformly over all strategies.

\medskip 

It remains to show sharpness. We first construct a one-round strategy using an
asymmetric phase-covariant qubit cloner~\cite{RezakhaniEtAl05}. To specify the cloner, let
\begin{align}
  \ket{0_Y}:=\frac{\ket0+i\ket1}{\sqrt2},
  \qquad
  \ket{1_Y}:=\frac{\ket0-i\ket1}{\sqrt2}
\end{align}
be the two eigenstates of the Pauli-$Y$ operator, and define the isometry
$V:A'\longrightarrow B\otimes C$ by
\begin{equation}
    V\ket{0_Y} =\ket{0_Y0_Y}_{BC},\qquad 
    V\ket{1_Y} =\eta_B\ket{1_Y0_Y}_{BC}
                  +\eta_C\ket{0_Y1_Y}_{BC},
  \label{eq:phase-covariant-cloner-isometry}
\end{equation}
where \begin{equation}
  \eta_B,\eta_C\geq0,
  \qquad \eta_B^2+\eta_C^2=1.
  \label{eq:cloning-circle}
\end{equation}
The strategy starts by preparing
\begin{equation}
  \ket{\Psi}_{ABC}
  :=(I_A\otimes V)\ket{\Phi}_{AA'},
  \qquad
  \ket{\Phi}_{AA'}:=\frac{\ket{00}+\ket{11}}{\sqrt2},
  \label{eq:clone-game-state}
\end{equation}
and distributing it among Alice, Bob and Charlie. 
Alice's outcome $x$ is uniform in either announced basis $\theta$.
Moreover, since the BB84 basis vectors are real in the computational
basis,
\begin{equation}
  (\bra{x^\theta}_A\otimes I_{BC})\ket{\Psi}_{ABC}
  =\frac1{\sqrt2}V\ket{x^\theta}.
  \label{eq:clone-conditional-state}
\end{equation}
Thus, conditioned on $x$ and $\theta$, Bob and Charlie share the normalized
state $V\ket{x^\theta}$. They measure their qubits in the announced BB84 basis $\theta$ and
output measurement outcomes $y, z$.

To compute the error probabilities, for
$x,\theta\in\{0,1\}$, define
\begin{equation}
  \ket{\phi_x^\theta}
  :=\frac{\ket{0_Y}+i^\theta(-1)^x \ket{1_Y}}{\sqrt2}.
  \label{eq:bb84-in-y-basis}
\end{equation}
These states represent the same measurement bases as
$\ket{x^\theta}=H^\theta\ket x$ up to global phases:
\begin{align}
  \ket{\phi_0^0}=\ket0,\quad
  \ket{\phi_1^0}=i\ket1,\quad
  \ket{\phi_0^1}=e^{i\pi/4}\ket+,\quad
  \ket{\phi_1^1}=e^{-i\pi/4}\ket-.
\end{align}
In particular, $V\ket{\phi_x^\theta}$ is the same as $V\ket{x^\theta}$ up to a global phase.
Thus, substituting~\eqref{eq:bb84-in-y-basis} into
\eqref{eq:phase-covariant-cloner-isometry} gives the measurement amplitude
\begin{equation}
  (\bra{\phi_y^\theta}\otimes\bra{\phi_z^\theta})
       V\ket{\phi_x^\theta}
  =\frac{1+(-1)^{x+y}\eta_B+(-1)^{x+z}\eta_C}{2\sqrt2}.
  \label{eq:clone-measurement-amplitude}
\end{equation}
For the error bits $E_B=x\oplus y$ and $E_C=x\oplus z$, this yields
\begin{align}
  &\Pr(E_B=e_B,E_C=e_C\mid x,\theta)\\
  &\quad=\frac18
     \left(1+(-1)^{e_B}\eta_B+(-1)^{e_C}\eta_C\right)^2\\
  &\quad=\frac14\left(1+(-1)^{e_B}\eta_B+(-1)^{e_C}\eta_C
                  +(-1)^{e_B+e_C}\eta_B\eta_C\right)\\
  &\quad=\frac14
     \left(1+(-1)^{e_B}\eta_B\right)
     \left(1+(-1)^{e_C}\eta_C\right),
  \label{eq:clone-joint-error-law}
\end{align}
where in the second equality we use $\eta_B^2+\eta_C^2=1$. The final expression does
not depend on $x$ or $\theta$, so it also gives the unconditional joint
error law. In particular, the two error bits are independent, with
\begin{equation}
  p_B:=\Pr(E_B=1)=\frac{1-\eta_B}{2},
  \qquad
  p_C:=\Pr(E_C=1)=\frac{1-\eta_C}{2}.
  \label{eq:clone-error-rates}
\end{equation}

Repeat this strategy independently on all $n$ coordinates. The total
error counts $N_B=d_H(x,y)$ and $N_C=d_H(x,z)$ are independent binomial
random variables with parameters $(n,p_B)$ and $(n,p_C)$, respectively.
Consequently, the optimal winning probability satisfies the explicit
lower bound
  \begin{align}
    p_{\mathrm{win}}^{(n)}(\beta, \gamma)
    &\geq\Pr(N_B\leq \beta n )
           \Pr(N_C\leq \gamma n)\\
    &=\left(\sum_{j=0}^{\lfloor\beta n\rfloor}
             \binom nj p_B^j(1-p_B)^{n-j}\right)
      \left(\sum_{k=0}^{\lfloor\gamma n\rfloor}
             \binom nk p_C^k(1-p_C)^{n-k}\right).   \label{eq:clone-binomial-winning-probability}
  \end{align}

If the reverse strict inequality to~\eqref{eq:two-sided-outside-circle}
holds, then $1-2\beta<\sqrt{1-(1-2\gamma)^2}$. In this case, choose $\eta_B$ strictly between these
two numbers and set $\eta_C=\sqrt{1-\eta_B^2}$. This gives
\begin{equation}
  \eta_B>1-2\beta,\qquad \eta_C>1-2\gamma,
  \qquad\implies \qquad p_B<\beta,\quad p_C<\gamma.
  \label{eq:interior-cloner-choice}
\end{equation}
Hoeffding's inequality and a union bound therefore imply
\begin{equation}
  p_{\mathrm{win}}^{(n)}(\beta,\gamma)
  \geq 1-e^{-2n(\beta-p_B)^2}-e^{-2n(\gamma-p_C)^2}.
  \label{eq:inside-circle-achievability}
\end{equation}
Both gaps are fixed and positive, proving that the winning probability
tends to $1$ exponentially fast.

On the boundary~\eqref{eq:two-sided-circle-boundary}, take
$\eta_B=1-2\beta$ and $\eta_C=1-2\gamma$, so that $p_B=\beta$ and
$p_C=\gamma$. Away from the two endpoints, both parameters lie in
$(0,1/2)$. For any fixed $p\in(0,1)$ and $N\sim\operatorname{Bin}(n,p)$,
the central limit theorem gives
\begin{equation}
  \Pr(N\leq\lfloor np\rfloor)
  =\Pr\left(
       \frac{N-np}{\sqrt{np(1-p)}}
       \leq\frac{\lfloor np\rfloor-np}{\sqrt{np(1-p)}}
      \right)
  \longrightarrow\frac12,
  \label{eq:binomial-boundary-limit}
\end{equation}
since the threshold on the right tends to zero. Applying this to both
factors in~\eqref{eq:clone-binomial-winning-probability} shows that the
constructed strategy has winning probability tending to $1/4$.
At $(\beta,\gamma)=(0,1/2)$, Bob makes no errors and Charlie's error count
has distribution $\operatorname{Bin}(n,1/2)$, so the winning probability
tends to $1/2$. The other endpoint is identical with the players
interchanged. Thus, at every boundary point,
\begin{equation}
  \liminf_{n\to\infty}
  p_{\mathrm{win}}^{(n)}(\beta,\gamma)\geq\frac14.
  \label{eq:boundary-achievability}
\end{equation}
\end{proof}

\section{Smooth min-entropy via a relaxation of quantum list decoding}
\label{sec:entropy-proof}

This section derives the smooth min-entropy estimate needed for the QKD
application. We keep Alice's BB84 measurements and Bob's bounded-error
condition, but replace the third player's guessing measurement by
positive test operators with a bound on their sum. This enlarges the
class of list-decoding strategies and yields a semidefinite optimization whose dual provides 
our desired smooth min-entropy bound. We then prove an overlap
estimate for the enlarged test class and apply the Boolean-cube bound
to obtain both a finite-block estimate and the asymptotic entropy rate.

Let $\rho_{ABC}$ be an arbitrary state, where $A$ consists of
$n$ qubits. Alice chooses $\Theta\in\{0,1\}^n$ uniformly,
independently of this state, and measures in the basis
$\{\ket{x^\theta}:x\in\{0,1\}^n\}$ given in~\eqref{eq:bb84-basis-vector}.
Her outcome is denoted by $X$. Bob learns the entire $\theta$ and applies
an arbitrary joint POVM $\{B_y^\theta\}_y$ to obtain $Y$.
Here $X,Y\in\{0,1\}^n$ denote the full random strings, with
realizations $x,y$. The third player retains the quantum system $C$.

For the entropy bound and its proof, fix $0\le\beta<1/2$.
For each block length $n$ and experiment as above, let
$\omega_{X\Theta C}$ be the subnormalized postmeasurement state restricted to
$d_H(X,Y)\le\beta n$, without dividing by this event's
probability.

For a subnormalized state $\tau_{XC}$, the conditional min-entropy is
defined by
\begin{equation}\label{eq:min-entropy-definition}
 H_{\min}(X|C)_\tau
 =-\log\inf\{\operatorname{Tr}\sigma_C:\sigma_C\ge0,\ \tau_{XC}\le I_X\otimes \sigma_C\}.
\end{equation}
Here $\sigma_C$ is a positive semidefinite operator on $C$ and need
not be normalized. We use purified-distance smoothing over subnormalized
states~\cite{Tomamichel16}:
\begin{equation}\label{eq:smooth-min-entropy-definition}
 H_{\min}^{\varepsilon}(X|C)_\tau
 =\sup_{\substack{\tau'\ge0,\ \operatorname{Tr}\tau'\le1\\P(\tau',\tau)\le\varepsilon}}
       H_{\min}(X|C)_{\tau'},
\end{equation}
where $P(\cdot, \cdot)$ denotes purified distance and $\operatorname{Tr}\tau'\le1$.
We allow the zero state as a smoothing witness and set $H_{\min}(X|C)_0=+\infty$.

\begin{proposition}[Smooth min-entropy bound]
\label{prop:entropy-rate}
For every $\nu>0$ and $\varepsilon\in(0,1)$, there is $n_0$,
depending only on $\beta,\nu,\varepsilon$, such that for all
$n\ge n_0$,
\begin{equation}
 H_{\min}^{\varepsilon}(X|C\Theta)_\omega
 \ge n\bigl(1-h_2(\beta)-\nu\bigr).
 \label{eq:main-entropy-rate}
\end{equation}
The bound is uniform over states and Bob's joint measurements resulting in $\omega_{X\Theta C}$ as defined above.
\end{proposition}

We prove this proposition by extending the operator-averaging method to
bounded-sum quantum tests.

\subsection{Quantum tests and smooth min-entropy}

For a subnormalized cq state
$\omega_{XC}=\sum_x\ketbra xx\otimes\omega_x$ and $L>0$, define
\begin{equation}
 \Ftest_L(\omega)=
 \max_{\substack{0\le M_x\le I_C\\\sum_xM_x\le LI_C}}
 \sum_x\operatorname{Tr}(\omega_xM_x).
 \label{eq:test-functional}
\end{equation}
If the conditioning system includes classical registers, the
tests can be pinched in those registers, so the sum constraint
holds in each classical block.

The above optimization is a relaxation of the list decoding problem. For integer $L\ge1$, a list decoder measures $C$ and outputs
a set $\mathcal L$ of at most $L$ candidates, succeeding when
$X\in\mathcal L$~\cite{Hayashi06}. A POVM
$\{N_{\mathcal L}\}_{\mathcal L}$ with such outcomes induces
\begin{equation}\label{eq:list-decoder-effects}
 M_x=\sum_{\mathcal L:\,x\in\mathcal L}N_{\mathcal L},
 \qquad
 \sum_xM_x=\sum_{\mathcal L}|\mathcal L|N_{\mathcal L}\le LI_C.
\end{equation}
Its success probability is therefore at most $\Ftest_L(\omega)$.
For classical side information, equality holds by selecting
the $L$ most likely candidates for each observed value.
For quantum side information, feasible effects need not arise
from a single measurement producing a list. Thus $\Ftest_L$
is a relaxation of list-decoding success; for $L=1$ it equals
the ordinary guessing probability.

Classically, deleting the $L$ most likely candidates for each
value of the side information leaves a subnormalized distribution
whose guessing probability is at most the total deleted mass
divided by $L$.
The following lemma gives a quantum counterpart, converting
a bound on the relaxed list-decoding success into smooth
min-entropy.

\begin{lemma}
\label{lem:test-smoothing}
Let $L>0$ and $\varepsilon\in(0,1)$. If
$\Ftest_L(\omega)\le\varepsilon^2/2$, then
\begin{equation}
 H_{\min}^{\varepsilon}(X|C)_\omega
 \ge\log L+\log\frac{2}{\varepsilon^2}.
 \label{eq:test-smoothing}
\end{equation}
\end{lemma}

\begin{proof}
Semidefinite duality gives
\begin{equation}
 \Ftest_L(\omega)=
 \min_{\sigma_C\ge0}\left\{L\operatorname{Tr} \sigma_C+
                   \sum_x\operatorname{Tr}(\omega_x-\sigma_C)_+\right\}.
 \label{eq:test-dual}
\end{equation}
Indeed, the dual constraints are $\sigma_C,U_x\ge0$ and
$\sigma_C+U_x\ge\omega_x$, and for fixed $\sigma_C$ the minimum of
$\operatorname{Tr} U_x$ is $\operatorname{Tr}(\omega_x-\sigma_C)_+$. Strict primal feasibility
follows by taking all $M_x$ to be sufficiently small positive
multiples of the identity, so the dual optimum is attained.

By assumption, an optimizer in~\eqref{eq:test-dual} satisfies
\begin{equation}\label{eq:dual-optimizer-bounds}
 \operatorname{Tr} \sigma_C\le\frac{\varepsilon^2}{2L},\qquad
 \operatorname{Tr}(\omega_{XC}-I_X\otimes \sigma_C)_+\le\frac{\varepsilon^2}{2}.
\end{equation}
If $\sigma_C=0$, then $\operatorname{Tr}\omega\le\varepsilon^2/2$ and the zero state is a valid smoothing witness, proving the claim. Otherwise, the standard positive-part smoothing bound
\cite[Lemma~6.21]{Tomamichel16}, applied to $I_X\otimes \sigma_C$,
gives a subnormalized $\widetilde\omega$ with
\begin{equation}\label{eq:smoothing-witness}
 \widetilde\omega\le I_X\otimes \sigma_C,\qquad
 P(\omega,\widetilde\omega)
 \le\sqrt{\varepsilon^2-\varepsilon^4/4}\le\varepsilon.
\end{equation}
Here the comparison operator can be normalized and its trace absorbed into the scale parameter of that lemma. If the smoothing radius is at least $\sqrt{\operatorname{Tr}\omega}$, the zero witness suffices instead. 
Then, by definition we have 
$H_{\min}(X|C)_{\widetilde{\omega}}\ge-\log\operatorname{Tr} \sigma_C
\ge\log L+\log(2/\varepsilon^2)$.
The witness $\widetilde \omega$ may be chosen to preserve the cq structure and
any classical conditioning registers
\cite[Lemma~6.13]{Tomamichel16}.
\end{proof}

\subsection{The BB84 overlap estimate}

To prove the min-entropy bound of Proposition~\ref{prop:entropy-rate},
we use Lemma~\ref{lem:test-smoothing}, which reduces the task
to bounding $\mathcal F_L(\omega)$.
We interpret this quantity as a relaxed success probability
and follow the operator-averaging approach of the previous section.
Indeed, the classical register $\Theta$ in $\omega_{X\Theta C}$
records Alice's measurement basis, and the state is already
restricted to the event $d_H(X,Y)\leq\beta n$ expressing Bob's
success. The effects defining $\mathcal F_L$ therefore take
the place of the third player's Hamming-ball test.
This gives an average of acceptance operators indexed by $\Theta$,
whose pairwise overlaps we estimate below.

After a common Naimark dilation, take Bob's $B_y^\theta$ to
be projectors. For each $\theta\in\{0,1\}^n$, define
\begin{equation}
 A_\theta=\sum_x\ketbra{x^\theta}{x^\theta}
  \otimes\left(\sum_{y:\,d_H(x,y)\le\beta n}B_y^\theta\right)
  \otimes M_x^\theta.
 \label{eq:acceptance-operator}
\end{equation}
The operators $A_\theta$ are
positive contractions and their average acceptance $\operatorname{Tr}[\rho_{ABC}\,2^{-n}\sum_\theta A_\theta]$ is equal to the objective in~\eqref{eq:test-functional} for
$\omega_{X\Theta C}$, because the projections on Bob's side postselect $\omega_{ X\Theta C}$ by imposing the condition $d_H(X, Y)\leq \beta n$.

\begin{lemma}
\label{lem:test-overlap}
Let $L>0$, and let $\{M_x^\theta\}_{x,\theta}$ be effects
satisfying
\begin{equation}
 0\le M_x^\theta\le I,\qquad
 \sum_xM_x^\theta\le LI
 \quad\text{for every }\theta.
 \label{eq:effect-budget}
\end{equation}
Then the operators $A_\theta$ defined
in~\eqref{eq:acceptance-operator} satisfy, for every
$\theta,\theta'\in\{0,1\}^n$,
\begin{equation}
 \opnorm{\sqrt{A_\theta}\sqrt{A_{\theta'}}}
 \le\sqrt{L\,2^{-d_H(\theta,\theta')}
       \sum_{j=0}^{\lfloor\beta n\rfloor}
       \binom{d_H(\theta,\theta')}{j}}.
 \label{eq:test-overlap}
\end{equation}
\end{lemma}

\begin{proof}
As in Lemma~\ref{lem:two-sided-overlap-reduction}, drop Bob's
condition in the first operator and the third player's condition in the
second:
\begin{align}
 P&=\sum_x\ketbra{x^\theta}{x^\theta}\otimes I_B\otimes M_x^\theta,\\
 Q&=\sum_yQ_y\otimes B_y^{\theta'}\otimes I_C,\qquad\qquad 
 Q_y=\sum_{x':\,d_H(x',y)\le\beta n}
             \ketbra{{x'}^{\theta'}}{{x'}^{\theta'}}.
\end{align}
Then $A_\theta\le P$, $A_{\theta'}\le Q$, and $Q$ is a
projector. For positive operators, the square-root overlap is monotone in each
argument: for example, $A\le P$ implies
$\|\sqrt A\sqrt B\|^2=\|\sqrt B A\sqrt B\|
\le\|\sqrt B P\sqrt B\|$. Applying this in both arguments gives
\begin{equation}
 \|\sqrt{A_\theta}\sqrt{A_{\theta'}}\|^2
 \le\|\sqrt P Q\sqrt P\|=\|QPQ\|.
 \label{eq:overlap-enlargement}
\end{equation}

Let $J=\{k:\theta_k\ne\theta'_k\}$, so $|J|=d_H(\theta,\theta')$.
We conjugate Alice's system by $\bigotimes_{k=1}^nH^{\theta_k}$.
This does not change the operator norm, and it sends the first basis
to the computational basis. The second basis becomes computational
on $J^c$ and Hadamard on $J$. We order Alice's tensor factors as
$J^c,J$ and continue to use $P$ and $Q$ for the conjugated operators.
Writing $x'=(w,v)$ in this order gives
\begin{equation}
 Q_y=\sum_{\substack{w\in\{0,1\}^{|J^c|},\ v\in\{0,1\}^{|J|}:\\
 d_H(w,y_{J^c})+d_H(v,y_J)\le\beta n}}
 \ketbra ww\otimes\ketbra{\widehat v}{\widehat v}.
 \label{eq:hadamard-ball-coordinate-split}
\end{equation}
where $\ket{\widehat v} = H^{\otimes |J|}\ket v$. Fixing $w$, we can then write 
\begin{equation}\label{eq:restricted-hadamard-ball}
 Q_y=\sum_w\ketbra ww\otimes Q_{y,w}, \qquad \qquad  Q_{y,w}=
 \sum_{\substack{v\in\{0,1\}^{|J|}:\\
  d_H(v,y_J)\le\beta n-d_H(w,y_{J^c})}}
 \ketbra{\widehat v}{\widehat v}.
\end{equation}

Since the operators $B_y^{\theta'}$ are mutually orthogonal projectors and
$P$ acts as the identity on Bob's system, $QPQ$ is block diagonal
in $y$. Within each such block, both $P$ and $Q$ are also block
diagonal in the computational string $w$ on $J^c$ as discussed above.
Therefore, the corresponding $(y,w)$ block is
\begin{equation}
 \sum_{u\in\{0,1\}^{|J|}}
 Q_{y,w}\ketbra uu Q_{y,w}\otimes M_{(w,u)}^\theta,
 \label{eq:test-overlap-block}
\end{equation}
where $x=(w,u)$ denotes the full string with restrictions $w$ on
$J^c$ and $u$ on $J$. Consequently, $\|QPQ\|$ is bounded by the maximum of the norms of these $(y,w)$ blocks.  

The rank of $Q_{y,w}$ is the number of strings $v$ in its defining
sum, so it is bounded by 
$\sum_{j=0}^{\lfloor\beta n\rfloor}\binom{|J|}{j}$.
Moreover, every computational vector $\ket u$ has squared overlap
$2^{-|J|}$ with every Hadamard vector in that sum. Hence
\begin{equation}
 \big\|Q_{y,w}\ket u\big\|^2
 =\bra u Q_{y,w}\ket u
 =2^{-|J|}\operatorname{rank}(Q_{y,w}).
 \label{eq:flat-overlap-projected-norm}
\end{equation}
The positive operator $Q_{y,w}\ketbra uu Q_{y,w}$, of rank at most one,
is supported on the range of $Q_{y,w}$, and its only possible
nonzero eigenvalue is $\|Q_{y,w}\ket u\|^2$.
It is therefore bounded above by this scalar times $Q_{y,w}$.
Combining this observation with the rank estimate gives, for every $u$,
\begin{equation}\label{eq:flat-overlap-rank-bound}
 Q_{y,w}\ketbra uu Q_{y,w}
 \le 2^{-|J|}
 \left(\sum_{j=0}^{\lfloor\beta n\rfloor}\binom{|J|}{j}\right)Q_{y,w}.
\end{equation}
Consequently, the corresponding block of $QPQ$ obeys
\begin{align}
 \sum_uQ_{y,w}\ketbra uu Q_{y,w}\otimes M_{(w,u)}^\theta
 &\le2^{-|J|}
       \left(\sum_{j=0}^{\lfloor\beta n\rfloor}\binom{|J|}{j}\right)Q_{y,w}
       \otimes\sum_uM_{(w,u)}^\theta\\
 &\le L\,2^{-|J|}
       \left(\sum_{j=0}^{\lfloor\beta n\rfloor}\binom{|J|}{j}\right)
       Q_{y,w}\otimes I_C.
\end{align}
For the last step, positivity and
\eqref{eq:effect-budget} give
$\sum_uM_{(w,u)}^\theta\le\sum_xM_x^\theta\le LI_C$
for each fixed $w$. Since $Q_{y,w}$ is a projector, the norm of
every block is at most
$L\,2^{-|J|}\sum_{j=0}^{\lfloor\beta n\rfloor}\binom{|J|}{j}$.
The same bound holds for $\|QPQ\|$, so
using~\eqref{eq:overlap-enlargement} and taking square roots proves
the claim. Empty restricted balls give zero blocks, while full
balls and $J=\varnothing$ are covered by the same argument with
the usual conventions.
\end{proof}

\subsection{From operator averages to smooth min-entropy}

Combining Lemma~\ref{lem:test-overlap} with
Theorem~\ref{thm:boolean-cube-bound} gives, for every $L>0$ and
integer $3\le D\le n$,
\begin{equation}
 \Ftest_L(\omega_{X\Theta C})
 \le 2^{-cD^2/n}
 +\sqrt{L\,2^{-(n-D)}
       \left(\sum_{j=0}^{D}\binom nj\right)
       \left(\sum_{k=0}^{\lfloor\beta n\rfloor}\binom{n-D}{k}\right)}.
 \label{eq:finite-test-bound}
\end{equation}
Indeed, the operators $A_\theta$ are contractions, and the binomial
tail $2^{-d}\sum_{k=0}^{\lfloor\beta n\rfloor}\binom{d}{k}$ is
nonincreasing in $d$. Thus the largest overlap needed
in~\eqref{eq:boolean-cube-final-bound} is bounded by the expression
from Lemma~\ref{lem:test-overlap} at $d=n-D$. The bound is uniform
over the state and the measurement strategy. It also holds for any
positive cq substate of $\omega_{X\Theta C}$, since the objective
in~\eqref{eq:test-functional} is positive.

\begin{proof}[Proof of Proposition~\ref{prop:entropy-rate}]
Set $L=2^{n(1-h_2(\beta)-\nu)}$ and $D=\lceil n^{2/3}\rceil$.
For every integer $0\le k\le n$, the standard Hamming-ball bound gives
\begin{equation}\label{eq:hamming-ball-cardinality}
 \sum_{j=0}^{\lfloor\beta n\rfloor}\binom kj
 \le \sum_{j=0}^{\lfloor\beta n\rfloor}\binom nj
 \le 2^{n h_2(\beta)}.
\end{equation}
Thus Lemma~\ref{lem:test-overlap} bounds the overlap by
$2^{(D-\nu n)/2}$ whenever $d_H(\theta,\theta')\ge n-D$.
Since $\sum_{j=0}^D\binom nj=2^{o(n)}$, the Boolean-cube
bound~\eqref{eq:boolean-cube-final-bound} and $\|A_\theta\|\le1$ yield
\begin{equation}
 \Ftest_L(\omega)
 \le \max_{\{M_x^\theta\}}
       \left\|2^{-n}\sum_\theta A_\theta\right\|
 \le 2^{-c n^{1/3}}+2^{-\nu n/2+o(n)}
 \longrightarrow0,
 \label{eq:asymptotic-test-bound}
\end{equation}
uniformly over states and measurements.
For all sufficiently large $n$, this is at most
$\varepsilon^2/2$. Lemma~\ref{lem:test-smoothing} therefore gives
\begin{equation}
 H_{\min}^{\varepsilon}(X|C\Theta)_\omega
 \ge \log L+\log\frac{2}{\varepsilon^2}
 \ge \log L
 = n(1-h_2(\beta)-\nu),
\end{equation}
which proves the proposition.
\end{proof}

\section{One-sided device-independent QKD}
\label{sec:qkd-proof}

We now apply the entropy estimates of Section~\ref{sec:entropy-proof}
to one-sided device-independent QKD. After specifying the protocol and
its secrecy criterion, we use random sampling to relate the observed
test error to the error rate of the full block. The finite-block bound
then follows by accounting for the public reconciliation data and
applying privacy amplification.

We consider the entanglement-based BB84 protocol E-QKD
from~\cite[Section~IV and Fig.~1]{TFKW13} using the same notation:
$n$ is the raw block length, $T\subseteq[n]$ is an independent uniformly
chosen test sample of size $t$, $\gamma$ is the tolerated sample
error rate, $S$ is an $s$-bit reconciliation syndrome, and $\ell$
is the final key length. Eve holds $E$, corresponding to the third player's system $C$ in
Section~\ref{sec:entropy-proof}.

Alice's quantum operations are trusted; Bob's measurement
device may behave arbitrarily and may make a joint measurement depending on the full basis string.

\begin{enumerate}
 \item \textbf{Preparation and confirmation.}
 Alice prepares $n$ EPR pairs $(\ket{00}+\ket{11})/\sqrt2$,
 keeping one qubit per pair and sending the others to Bob.
 Bob confirms receipt; his device must then remain isolated
 from Eve.
 \item \textbf{Measurement.}
 Alice announces uniform $\Theta\in\{0,1\}^n$ and measures in
 those BB84 bases to obtain $X$. Bob's device produces $Y$
 using any joint $\Theta$-dependent POVM.
 \item \textbf{Parameter estimation.}
 After $Y$ is fixed, Alice chooses an independent uniform
 $t$-element sample $T\subseteq [n]$ and announces $T,X_T$.
 They abort and set $K=\perp$ unless 
 \begin{equation}
d_H(X_T, Y_T)\le\gamma t.
 \end{equation}
 \item \textbf{Reconciliation and privacy amplification.}
 On acceptance, Alice sends the $s$-bit syndrome
 $S=\operatorname{syn}(X_{T^c})$, where $\operatorname{syn}$ is a fixed reconciliation map, and the description of a hash function $F: \{0,1\}^{n-t}\to \{0,1\}^\ell$, chosen independently from a two-universal family.
 She outputs $K=F(X_{T^c})$.
 
 \item \textbf{Bob's output key.} Using the syndrome $S$, Bob applies the reconciliation procedure to $Y_{T^c}$
to obtain an estimate $\widehat{X}_{T^c}$ of Alice's remaining
raw string $X_{T^c}$. He then applies the same hash function $F$
to produce his key $\widehat K=F(\widehat{X}_{T^c})$.
\end{enumerate}

The bit-error rate between Alice's and Bob's $n$-bit outcome
strings $X$ and $Y$ is $d_H(X,Y)/n$, the fraction of positions
at which their bits disagree. The protocol estimates this rate
using the random test sample $T$ and accepts only if the observed
sample error rate is at most a prescribed threshold $\gamma$.

As in~\cite{TFKW13}, reconciliation enters the secrecy bound only
through its leakage $s$. We do not reanalyse its correctness or the
probability of an honest abort. A complete QKD implementation must
combine the secrecy bound below with the corresponding correctness
guarantee for reconciliation.

Write $E^+=(E,\Theta,T,X_T,S,F)$ for Eve's final information.
We call the protocol $\delta$-secret, in the sense of
\cite[Eq.~(10)]{TFKW13}, if
\begin{equation}
 \frac{\Pr[K\ne\perp]}{2}\,
 \left\|\rho_{KE^+|K\ne\perp}
       -\mu_K\otimes\rho_{E^+|K\ne\perp}\right\|_1
 \le\delta,
 \label{eq:security-definition}
\end{equation}
where $\|\cdot\|_1$ is the trace norm and $\mu_K=2^{-\ell}I_K$ corresponds to a uniform $\ell$-bit key. Moreover, $\rho_{KE^+|K\ne\perp}$ is the normalized postselected state conditioned on $K\neq \perp$. In particular, the left side is zero if the protocol
always aborts.

\begin{theorem}[Asymptotic secrecy of E-QKD]
\label{thm:qkd}
Let the tolerated bit-error rate $\gamma$ satisfy $0\le\gamma<1/2$. In the asymptotically optimal
reconciliation regime $s/n\to h_2(\gamma)$ every rate
\begin{equation}
 0<R<1-2h_2(\gamma)
 \label{eq:qkd-rate}
\end{equation}
can be certified. That is, the parameters of E-QKD can be chosen so that
$\ell/n\to R$ and the secrecy error $\delta$ tends to zero
as $n\to\infty$, uniformly over Bob's measurement devices.
\end{theorem}

We note that the positive-rate condition is $h_2(\gamma)<1/2$,
corresponding to a noise threshold of approximately $11\%$, same as in the usual device-dependent BBM92 protocol.

\subsection{Finite-block secrecy}

We follow the sampling, reconciliation, and privacy-amplification
argument of~\cite[Theorem~5]{TFKW13}, using the smooth-entropy
bound from Section~\ref{sec:entropy-proof}. In the following we allow any
initial state $\rho_{ABE}$ with $A$ consisting of $n$ qubits,
independent of $\Theta,T$. The constant $c>0$ below is the universal
constant in Theorem~\ref{thm:boolean-cube-bound}.

\begin{proposition}[Finite-block secrecy bound]
\label{prop:qkd-finite}
Consider E-QKD
with $n\ge3$, $1\le t<n$, $0\le\gamma<1/2$, syndrome length
$s$, and key length $\ell$. Let $\varepsilon>0$ with
$\gamma+\varepsilon<1/2$, and let $\varepsilon'\in(0,1)$ be the
smoothing parameter. Choose $L>0$ and an integer $3\le D\le n$
such that
\begin{equation}
 2^{-cD^2/n}
   +\sqrt{L\,2^{-(n-D)}
      \left(\sum_{j=0}^D\binom nj\right)
      \left(\sum_{k=0}^{\lfloor(\gamma+\varepsilon)n\rfloor}
             \binom{n-D}{k}\right)}
 \le\frac{\varepsilon'^2}{2}.
 \label{eq:qkd-error}
\end{equation}
Then the protocol is $\delta$-secret with
\begin{equation}
 \delta=e^{-2\varepsilon^2t}
       +2\varepsilon'
       +\frac{\varepsilon'}2\sqrt{\frac{2^{\ell+t+s}}{2L}}.
 \label{eq:qkd-security}
\end{equation}
\end{proposition}

\begin{proof}

Let $\rho_{\Theta T XYE}$ be the corresponding postmeasurement
state, with $T$ still independent of the raw experiment.
As in the proof of~\cite[Theorem~5]{TFKW13}, let $\Omega$ be the event
\begin{equation}
\frac{d_H(X,Y)}{n}
        \le \frac{d_H(X_T,Y_T)}{t}+\varepsilon,
 \label{eq:sampling-event}
 \end{equation}
and $\Gamma$ be the event
 \begin{equation}
    d_H(X_T, Y_T)\le\gamma t.
  \label{eq:sample-test}
 \end{equation} 
 Hoeffding's inequality for sampling without replacement,
applied conditionally on $X,Y$ and then averaged over these
strings, gives~\cite{TFKW13}
\begin{equation}
 \Pr[\Omega^c]\le e^{-2\varepsilon^2t}.
 \label{eq:sampling-error}
\end{equation}
Retain the subnormalized state
\begin{equation}
 \widetilde\rho_{\Theta T XYE}
 =\Pr[\Omega\cap\Gamma]\,
   \rho_{\Theta T XYE|\Omega\cap\Gamma}.
 \label{eq:qkd-good-state}
\end{equation}
This is the subnormalized accepted branch further restricted to $\Omega$.
The discarded accepted mass is
$\Pr[\Gamma\cap\Omega^c]\le\Pr[\Omega^c]\le e^{-2\varepsilon^2t}$.
This is an unconditional probability bound; we do not divide by the
acceptance probability.

On $\Omega\cap\Gamma$ we have
$d_H(X,Y)\le(\gamma+\varepsilon)n$.
Include the independent register $T$ with Eve's system and apply
the finite test bound~\eqref{eq:finite-test-bound}, with $C$ replaced
by $ET$ and $\beta=\gamma+\varepsilon$. For the state restricted to
$d_H(X,Y)\le(\gamma+\varepsilon)n$, the left-hand side
of~\eqref{eq:qkd-error} bounds $\Ftest_L$ from above and is at most $\varepsilon'^2/2$.
The objective in~\eqref{eq:test-functional} is positive, so
the same bound holds for its cq substate
$\widetilde\rho_{X\Theta TE}$ and we have
$\Ftest_L(\widetilde\rho_{X\Theta TE})\le \varepsilon'^2/2$.
Lemma~\ref{lem:test-smoothing} now gives
\begin{equation}
 H_{\min}^{\varepsilon'}(X|\Theta TE)_{\widetilde\rho}
 \ge\log L+\log\frac{2}{\varepsilon'^2}.
 \label{eq:qkd-raw-entropy}
\end{equation}
The classical leakage chain rule
\cite[Lemma~6.18]{Tomamichel16}, exactly as in
\cite[Eq.~(12)]{TFKW13}, then yields
\begin{align}
 H_{\min}^{\varepsilon'}(X_{T^c}|\Theta T X_T S E)_{\widetilde\rho}
 &\ge H_{\min}^{\varepsilon'}(X|\Theta TE)_{\widetilde\rho}-t-s\\
 &\ge\log L+\log\frac{2}{\varepsilon'^2}-t-s.
 \label{eq:qkd-chain-rule}
\end{align}
Here $X_T$ and $S$ are copied from $X$ as prescribed by the
protocol. Once $T,X_T$ are given, $X$ and $X_{T^c}$ carry
the same unknown information. The chain rule preserves the
smoothing parameter.

Smooth privacy amplification with two-universal hashing
\cite{TSSR11,Tomamichel16} now gives, on this subnormalized
branch,
\begin{equation}
 \frac12\big\|
 \widetilde\rho_{KE^+}
 -\mu_K\otimes\widetilde\rho_{E^+}\big\|_1
 \le2\varepsilon'
     +\frac{\varepsilon'}2\sqrt{\frac{2^{\ell+t+s}}{2L}}.
 \label{eq:qkd-hashing}
\end{equation}
To return to the actual accepted branch, write
$\rho^{\mathrm{acc}}_{KE^+}:=\Pr[\Gamma]\rho_{KE^+|\Gamma}$.
Reconciliation and hashing preserve positivity and trace, so
$\Delta_{KE^+}:=\rho^{\mathrm{acc}}_{KE^+}-\widetilde\rho_{KE^+}$
is positive and has trace at most $e^{-2\varepsilon^2t}$.
Its marginal $\Delta_{E^+}$ has the same trace. The triangle inequality gives
\begin{align}
 \frac12\big\|\rho^{\mathrm{acc}}_{KE^+}
       -\mu_K\otimes\rho^{\mathrm{acc}}_{E^+}\big\|_1
 &\le \frac12\big\|\widetilde\rho_{KE^+}
       -\mu_K\otimes\widetilde\rho_{E^+}\big\|_1
       +\frac12\big\|\Delta_{KE^+}
       -\mu_K\otimes\Delta_{E^+}\big\|_1\\
 &\le \frac12\big\|\widetilde\rho_{KE^+}
       -\mu_K\otimes\widetilde\rho_{E^+}\big\|_1
       +\operatorname{Tr}\Delta_{KE^+}.
 \label{eq:qkd-restore-accepted-branch}
\end{align}
The left-hand side equals the secrecy criterion in
\eqref{eq:security-definition}. Combining this inequality with
\eqref{eq:qkd-hashing} proves~\eqref{eq:qkd-security}, without
conditioning the sampling-error bound on acceptance.

\end{proof}

\subsection{Asymptotic key rate}

\begin{proof}[Proof of Theorem~\ref{thm:qkd}]
Fix a rate satisfying
\eqref{eq:qkd-rate} and choose $r$ with
$R+h_2(\gamma)<r<1-h_2(\gamma)$.
Take $\ell=\lfloor Rn\rfloor$, $L=2^{rn}$, $D=\lceil n^{2/3}\rceil$,
$t=\lceil n^{2/3}\rceil$, and $\varepsilon=\varepsilon'=n^{-1/6}$.
For all sufficiently large $n$, $\gamma+\varepsilon<1/2$.
By continuity and the strict gap in $r$, the binomial estimates
in the proof of Proposition~\ref{prop:entropy-rate} show that the
left-hand side of~\eqref{eq:qkd-error} is at most
$2^{-c_0 n^{1/3}}$ for some constant $c_0>0$ and all sufficiently
large $n$. Thus~\eqref{eq:qkd-error} holds for all sufficiently large $n$.
Also, $\varepsilon'\to0$,
$\varepsilon^2t\to\infty$ and, under $s/n\to h_2(\gamma)$,
\begin{equation}\label{eq:asymptotic-key-leakage}
 \frac{\ell+t+s}{n}\longrightarrow R+h_2(\gamma)<r.
\end{equation}
Each term of~\eqref{eq:qkd-security} therefore tends to zero,
uniformly over Bob's measurement devices, proving the theorem.
\end{proof}

\paragraph*{Acknowledgements.}
M.T. acknowledges support from the National Research Foundation Investigatorship Award (NRF-NRFI10-2024-0006) and the National Research Foundation, Singapore, through the National Quantum Office, hosted by A*STAR, under its Centre for Quantum Technologies Funding Initiative (S24Q2d0009).

\paragraph*{Use of artificial intelligence.} OpenAI Codex with ChatGPT~6 Astra assisted with exploring and checking proof strategies, and with preparing the manuscript. The current exposition of the proofs is due to the authors. The authors are responsible for the mathematical claims, the final presentation, and any remaining errors.


\appendix

\section{Complementary Hamming-ball overlap}
\label{app:elementary-ball-overlap}

This appendix gives a direct proof of the complementary Hamming-ball
overlap estimate used in Section~\ref{sec:robust-bb84}. It establishes
the required special case of Lemma~\ref{lem:Polyanskiy-Samorodnitsky}
with an explicit exponential rate for every positive block length.
The proof bounds the two projectors by exponential functions of
collective Pauli observables and reduces the resulting norm estimate
to a single-qubit calculation.

\begin{lemma}
  \label{lem:elementary-ball-overlap}
  Let $\beta,\gamma\in[0,1/2]$ satisfy
  \begin{equation}
    (1-2\beta)^2+(1-2\gamma)^2>1,
    \label{eq:appendix-ball-condition}
  \end{equation}
  and define
  \begin{equation}
    \label{eq:explicit-ball-exponent}
    \varepsilon_{\mathrm{ball}}(\beta,\gamma)
    :=
    \frac{1}{4\ln 2}
    \left(
      \sqrt{(1-2\beta)^2+(1-2\gamma)^2}-1
    \right)^2
    >0.
  \end{equation}
  Then, for every positive integer $n$,
  \begin{equation}
    \label{eq:explicit-ball-overlap}
    \opnorm{
      \Pi^{(n)}_{\gamma n}
      \widehat\Pi^{(n)}_{\beta n}
    }
    \leq
    2^{-n\varepsilon_{\mathrm{ball}}(\beta,\gamma)}.
  \end{equation}
\end{lemma}

\begin{proof}
  Let $Z$ and $X$ be the Pauli matrices and define the collective observables
  \begin{align}
    Z^{(n)}:=\sum_{j=1}^n
    I^{\otimes(j-1)}\otimes Z\otimes I^{\otimes(n-j)}, \label{eq:collective-z-observable}\\
    X^{(n)}:=\sum_{j=1}^n
    I^{\otimes(j-1)}\otimes X\otimes I^{\otimes(n-j)}. \label{eq:collective-x-observable}
  \end{align}
  A computational-basis vector of weight $k$ has $Z^{(n)}$-eigenvalue $n-2k$,
  and the analogous statement holds for $X^{(n)}$ in the Hadamard basis. Hence
  \begin{equation}
    \label{eq:ball-spectral-thresholds}
    (1-2\gamma)n\Pi^{(n)}_{\gamma n}\leq Z^{(n)}\Pi^{(n)}_{\gamma n},
    \qquad
    (1-2\beta)n\widehat\Pi^{(n)}_{\beta n}\leq X^{(n)}\widehat\Pi^{(n)}_{\beta n}.
  \end{equation}
  It follows that, for every $u,v\geq0$,
  \begin{equation}
    \label{eq:z-exponential-filter}
    \opnorm{\Pi^{(n)}_{\gamma n}e^{-uZ^{(n)}/2}}
    \leq e^{-u(1-2\gamma)n/2},
  \end{equation}
  \begin{equation}
    \label{eq:x-exponential-filter}
    \opnorm{e^{-vX^{(n)}/2}\widehat\Pi^{(n)}_{\beta n}}
    \leq e^{-v(1-2\beta)n/2}.
  \end{equation}
  Factoring the product of the two projectors and using submultiplicativity of the operator norm give
  \begin{align}
    \opnorm{\Pi^{(n)}_{\gamma n}\widehat\Pi^{(n)}_{\beta n}}
    & =  \opnorm{\big( \Pi^{(n)}_{\gamma n} e^{-uZ^{(n)}/2}\big)\big(e^{uZ^{(n)}/2}e^{vX^{(n)}/2}\big)\big(e^{-vX^{(n)}/2}\widehat\Pi^{(n)}_{\beta n}\big)}\\
    &\leq
    e^{-\frac n2[u(1-2\gamma)+v(1-2\beta)]}
    \opnorm{e^{uZ^{(n)}/2}e^{vX^{(n)}/2}} \\
    &=
    e^{-\frac n2[u(1-2\gamma)+v(1-2\beta)]}
    \opnorm{e^{uZ/2}e^{vX/2}}^n,
    \label{eq:ball-overlap-filter-bound}
  \end{align}
  where the last equality follows by tensorization.

Set $K_{u,v}:=e^{uZ/2}e^{vX/2}$. Multiplicativity of the determinant gives 
   \begin{equation}
    \det(K_{u,v}^\dagger K_{u,v})=\det\big(e^{uZ}\big)\det\big(e^{vX}\big)=1.
   \end{equation}
We also have    
   \begin{equation}	
    \frac12\operatorname{Tr}(K_{u,v}^\dagger K_{u,v}) = \frac12\operatorname{Tr}\big(e^{uZ}e^{vX}\big) = \frac 12\Big(e^{u}\bra{0} e^{vX} \ket{0}+ e^{-u}\bra{1} e^{vX} \ket {1}\Big) 
    =\cosh u\cosh v.
    \label{eq:single-qubit-filter-data}
  \end{equation}
  A positive $2\times 2$ matrix of determinant $1$ and half-trace $c\geq1$
  has largest eigenvalue $c+\sqrt{c^2-1}$. Consequently,
  \begin{equation}
    \label{eq:single-qubit-filter-norm}
    \opnorm{K_{u,v}}^2
    =c+\sqrt{c^2-1},
    \qquad c:=\cosh u\cosh v.
  \end{equation}
  Using $\cosh x\leq e^{x^2/2}$ we find that $c\leq e^t$ for
  $t:=(u^2+v^2)/2$. Hence
  \begin{align}
    \opnorm{K_{u,v}}^2& = c+\sqrt{c^2-1}\\
    &\leq e^t+\sqrt{e^{2t}-1}
     =e^t\left(1+\sqrt{1-e^{-2t}}\right) \\
    &\leq e^t(1+\sqrt{2t})
     \leq e^{t+\sqrt{2t}}.
    \label{eq:single-qubit-filter-elementary-bound}
  \end{align}

  Let
  \begin{equation}
    \label{eq:appendix-r}
    r:=\sqrt{(1-2\beta)^2+(1-2\gamma)^2}>1,
  \end{equation}
  and choose
  \begin{equation}
    u=\frac{r-1}{r}(1-2\gamma),
    \qquad
    v=\frac{r-1}{r}(1-2\beta).
  \end{equation}
  Then $t=\frac 12(u^2+v^2)=\frac 12 (r-1)^2$ and
  \begin{equation}
    \opnorm{K_{u,v}}
    \leq
    \exp\!\left(\frac{1}{4}(r-1)^2+\frac{1}{2}(r-1)\right).
    \label{eq:single-qubit-filter-radial-bound}
  \end{equation}
  Substitution into \eqref{eq:ball-overlap-filter-bound} yields
  \begin{align}
    \opnorm{\Pi^{(n)}_{\gamma n}\widehat\Pi^{(n)}_{\beta n}}
    & \leq e^{-\frac n2[u(1-2\gamma)+v(1-2\beta)]}
    \opnorm{K_{u, v}}^n\\
    & \leq \exp\left[ -\frac n2\left(  r(r-1) -\frac{1}{2}(r-1)^2 - (r-1)  \right)  \right]\\
    &=\exp\!\left[-\frac n4(r-1)^2\right]
     =2^{-n\varepsilon_{\mathrm{ball}}(\beta,\gamma)},
  \end{align}
  where the last equality is \eqref{eq:explicit-ball-exponent}.
\end{proof}

\section{Subspace coset monogamy games}
\label{sec:grassmannian-coset}

This appendix gives a further application of
Theorem~\ref{thm:matrix-approximation} to monogamy games whose questions
are indexed by subspaces. The normalized adjacency matrix of the
disjointness relation provides an exponentially accurate approximation
to the averaging projection. We first establish the resulting operator
bound, then apply it to the subspace coset game using the pairwise-overlap
estimate of Culf and Vidick~\cite{CulfVidick22}, and compare the resulting
winning-probability bound with earlier estimates.

The subspace coset monogamy problem was introduced by Coladangelo,
Liu, Liu, and Zhandry~\cite{ColadangeloEtAl21} in their study of
unclonable cryptography. They conjectured a monogamy property for coset states against
information-theoretic adversaries and showed its relevance to
unclonable decryption and copy-protection. Culf and
Vidick~\cite{CulfVidick22} proved this conjecture by showing that the
winning probability decays exponentially with the number of qubits.

Let $q$ be a prime power and let $\mathcal G=\operatorname{Gr}_q(2n,n)$
denote the set of $n$-dimensional linear subspaces of $\mathbb F_q^{2n}$,
where $n\ge1$. For two subspaces $V, W\in \mathcal G$ define
\begin{equation}\label{eq:grassmann-distance}
 d_{\mathcal G}(V,W):=n-\dim(V\cap W).
\end{equation}
The largest distance between two subspaces is $n$, attained precisely when $V$ and $W$
are complementary subspaces with a trivial intersection.

\begin{lemma}
\label{lem:grassmannian-bound}
Let $\{A_V:V\in\mathcal G\}$ be positive semidefinite operators on a common Hilbert space. 
Then
\begin{equation}\label{eq:grassmannian-bound}
 \opnorm{\frac1{|\mathcal G|}\sum_{V\in\mathcal G}A_V}
 \le \, \frac 1{q^n}\max_{V\in\mathcal G}\opnorm{A_V}+\max_{V, W: V\cap W=\{0\}} \opnorm{\sqrt{A_V}\sqrt{A_W}}.
\end{equation}
\end{lemma}

\begin{proof}
Apply Theorem~\ref{thm:matrix-approximation} with the uniform distribution $p_V=1/|\mathcal G|$, so that every entry of $R$ equals $1/|\mathcal G|$, and choose
\begin{equation}\label{eq:grassmannian-approximation}
 \tilde{r}_{V,W}:=\frac 1{q^{n^2}}\mathbf 1_{\{V\cap W=\{0\}\}},
\end{equation}
so $ \tilde{r}_{V,W}$ is nonzero only if $V, W$ are complementary subspaces. 
Every $V\in \mathcal G$ has exactly $q^{n^2}$ complements. To verify this, extend a basis $\{e_1,\ldots,e_n\}$ of $V$ to a basis $\{e_1,\ldots,e_{2n}\}$ of $\mathbb F_q^{2n}$. Then, each complement of $V$ has a unique representation of the form 
\begin{align}
\text{span}\{e_i + v_i:\, n+1\leq i\leq 2n\},
\end{align}
for some vectors $v_{n+1}, \dots, v_{2n}\in V$.
Since there are $q^n$ choices for each $v_i$, the total number of complementary subspaces is $q^{n^2}$. Thus $\widetilde{R}$ is a symmetric stochastic matrix and commutes with $R$. To bound $\|R-\widetilde{R}\|$ we need to find the eigenvalues of $\widetilde{R}$. 

We note that, by definition, $\widetilde{R}$ is equal to the normalized
adjacency matrix of the $q$-Kneser graph, whose eigenvalues are computed by Lv and Wang~\cite[Theorem~7]{LvWang12}:
\begin{equation}\label{eq:grassmannian-spectrum}
 \lambda_j(\widetilde{R})=(-1)^j q^{-nj+\binom j2},\qquad 0\le j\le n.
\end{equation}
The eigenvalue $\lambda_0=1$ is simple and corresponds to the constant
vector. Since $R$ is the orthogonal projection onto this vector, we have
\begin{equation}\label{eq:grassmannian-spectral-error}
 \opnorm{R-\widetilde{R}}
 =\max_{1\le j\le n}q^{-nj+\binom j2}=q^{-n}.
\end{equation}

Next, to apply Theorem~\ref{thm:matrix-approximation}, we need to choose a complete orthogonal family of permutations. To this end, we think of a complete bipartite graph between two copies of
$\mathcal G$. Then, such a family of permutations corresponds to a decomposition of edges of this bipartite graph into perfect matchings. We think of this complete bipartite graph as a disjoint union of two graphs: one consisting of edges $(V, W)$ of complementary subspaces, and the other one consisting of the rest of the edges. We note that by the above computations, both of these bipartite subgraphs are regular, and any regular bipartite graph can be decomposed into perfect matchings. Thus, there is a complete orthogonal family of permutations $\mathcal P$ consisting of two types of permutations: $\mathcal P = \mathcal P_1\cup \mathcal P_2$ such that
\begin{align}
&V\cap \pi(V) = \{0\}, \quad  \forall V\in \mathcal G, \pi \in \mathcal P_1,\\
&V\cap \pi(V) \neq \{0\}, \quad  \forall V\in \mathcal G, \pi \in \mathcal P_2.
\end{align}
Moreover, $|\mathcal P_1|=q^{n^2}$ and $|\mathcal P_2|=|\mathcal G|-q^{n^2}$.
On the other hand, by the definition of $\widetilde{R}$, if $\pi\in \mathcal P_1$, then $\tilde{r}_{V, \pi(V)} = q^{-n^2}$ and if $\pi\in \mathcal P_2$, then $\tilde{r}_{V, \pi(V)}=0$. Therefore, 
\begin{align}
 \sum_{\pi}\max_V
 \left\{|\tilde{r}_{V,\pi(V)}|
 \opnorm{\sqrt{A_V}\sqrt{A_{\pi(V)}}}\right\} &=  \sum_{\pi\in \mathcal P_1}\max_V
 \left\{|\tilde{r}_{V,\pi(V)}|
 \opnorm{\sqrt{A_V}\sqrt{A_{\pi(V)}}}\right\}\\
& \le q^{n^2}q^{-n^2}\max_{V, W: V\cap W=\{0\}} \opnorm{\sqrt{A_V}\sqrt{A_W}}\\
& = \max_{V, W: V\cap W=\{0\}} \opnorm{\sqrt{A_V}\sqrt{A_W}}.\label{eq:grassmannian-permutation-contribution}
\end{align}
Combining \eqref{eq:grassmannian-spectral-error} and
\eqref{eq:grassmannian-permutation-contribution} with
Theorem~\ref{thm:matrix-approximation} proves the claim.
\end{proof}

We now specialize to $q=2$. For $V\in\mathcal G=\operatorname{Gr}_2(2n,n)$, let
$\operatorname{CS}(V)$ be a fixed set of coset representatives and let
$V^\perp$ be its orthogonal complement under the standard binary inner
product. Then, we define the coset states as
\begin{equation}\label{eq:coset-state-definition}
 |V_{x,z}\rangle
 :=\frac 1{2^{\frac n2}}\sum_{v\in V}(-1)^{z\cdot v}|x+v\rangle,
 \quad x\in\operatorname{CS}(V),\quad z\in\operatorname{CS}(V^\perp).
\end{equation}
For each $V$, these states form an orthonormal measurement basis for the space of $2n$ qubits. Based on this observation, the coset monogamy game is defined as follows. 
The referee chooses $V, x, z$ uniformly at random, prepares $|V_{x,z}\rangle$, and
sends it through a channel with outputs held by Bob and Charlie. After
$V$ is announced, Bob must identify $x+V$ and Charlie must identify $z+V^\perp$, without communicating. The optimal winning probability of this game is denoted by~$p_{\mathrm{coset}}^{(2n)}$.

\begin{corollary}\label{cor:coset-winning-bound}
For every $n\ge1$,
\begin{equation}\label{eq:coset-winning-bound}
 p_{\mathrm{coset}}^{(2n)}\le 2^{-n}+2^{-\frac n2}.
\end{equation}
\end{corollary}

\begin{proof}
Fix a strategy and, after local Naimark dilations, denote the projective measurements
of Bob and Charlie by $\{B_x^V:\, x \in \operatorname{CS}(V)\}_x$ and $\{C_z^V:\, z\in\operatorname{CS}(V^\perp)\}$, respectively. Define
\begin{equation}\label{eq:coset-acceptance-projection}
 A_V:=\sum_{x,z}|V_{x,z}\rangle\!\langle V_{x,z}|
       \otimes B_x^V\otimes C_z^V.
\end{equation}
Each $A_V$ is a projector and the winning probability of this strategy is
$|\mathcal G|^{-1}\sum_V\operatorname{Tr}(\rho_{ABC}A_V)\leq \opnorm{|\mathcal G|^{-1}\sum_V A_V}$.

The overlap estimate of Culf and Vidick~\cite[Lemma~3.2 and Eq.~(9)]{CulfVidick22} gives
\begin{equation}\label{eq:coset-acceptance-overlap}
 \opnorm{A_VA_W}
 \le 2^{-\frac12(n-\dim(V\cap W))}
 =2^{-d_{\mathcal G}(V,W)/2}.
\end{equation}
As in Lemma~\ref{lem:two-sided-overlap-reduction}, this follows by dropping
Bob's acceptance condition from one projection and Charlie's from the
other, and then decomposing their product into measurement-outcome blocks.
The remaining referee blocks are the coset projection products bounded
in~\cite[Lemma~3.2 and Eq.~(9)]{CulfVidick22}. This, in particular, implies that 
\begin{equation}
 \opnorm{A_VA_W}
 \le 2^{-\frac n2}, \qquad \forall V, W: V\cap W=\{0\}.
 \end{equation}
 Applying Lemma~\ref{lem:grassmannian-bound} and using $\opnorm{A_V}\le1$ proves
\eqref{eq:coset-winning-bound}.
\end{proof}

For comparison, Culf and Vidick~\cite[Theorem~2.1]{CulfVidick22} proved
\begin{equation}\label{eq:culf-vidick-comparison}
 p_{\mathrm{coset}}^{(2n)}
 \le\sqrt e\,\bigl(\cos(\pi/8)\bigr)^{2n}
 =\sqrt e\,2^{-\alpha n},
 \qquad \alpha:=-2\log\cos(\pi/8)\approx0.2284.
\end{equation}
Schleppy and Soljanin~\cite[Theorem~1 and Section~IV]{SchleppySoljanin25}
subsequently improved the exponent per qubit to $1/4$. More generally, for coset
states associated with $k$-dimensional subspaces of $\mathbb F_2^n$,
they obtain a winning-probability bound
$O(2^{-\min\{k,n-k\}/2})$.
In the balanced case, their proof gives the explicit estimate
\begin{equation}\label{eq:schleppy-soljanin-comparison}
 p_{\mathrm{coset}}^{(2n)}
 \le c 2^{-\frac n2}, \qquad  c= \left(\frac{7\sqrt 2 +9}{9-2\sqrt 2}\right)\approx 3.0623.
\end{equation}
Corollary~\ref{cor:coset-winning-bound} recovers the same exponent but with the prefactor $(1+2^{-\frac n2})<c$.
To compare the two proofs, the approximating matrix $\widetilde{R}$ allows us to retain only the overlaps of complementary subspaces and account for
the discarded relations through the spectral error of $\widetilde{R}$, while the bound of~\cite{SchleppySoljanin25} works directly with $R$. 

These estimates do not determine the optimal exponent of the game.
Sending the entire state to Bob lets him recover his coset perfectly,
while Charlie can guess his uniformly random coset label, giving
$p_{\mathrm{coset}}^{(2n)}\ge2^{-n}$. Thus a gap remains between the
exponent $1/2$ per qubit achieved by this strategy and the upper-bound exponent
$1/4$ per qubit. As discussed by Schleppy and Soljanin~\cite[Section~VI]{SchleppySoljanin25},
the pairwise-overlap approach reaches its smallest overlap estimate at
complementary subspaces. Improving the exponent would require additional
information about the acceptance projections beyond the estimates in
\eqref{eq:coset-acceptance-overlap}.

\end{document}